\documentclass[lettersize,journal]{IEEEtran}
\usepackage{amsmath,amsfonts}
\usepackage[ruled,linesnumbered]{algorithm2e}
\usepackage{array}
\usepackage{textcomp}
\usepackage{stfloats}
\usepackage{url}
\usepackage{verbatim}
\usepackage{graphicx}
\usepackage{cite}
\usepackage{color}
\usepackage[colorlinks,linkcolor = black,anchorcolor = black,citecolor = black, urlcolor=black]{hyperref}
\usepackage{subcaption}        
\usepackage{booktabs}

\usepackage{xspace}
\newcommand{\name}{DeepGuard\xspace}

\begin{document}
\title{\name: Defending Deep Joint Source-Channel Coding Against Eavesdropping at Physical-Layer}

\author{
Kaiyi Chi,
Yinghui He,
Qianqian Yang,~\IEEEmembership{Member,~IEEE},
Yuanchao Shu,~\IEEEmembership{Senior Member,~IEEE},

Zhiqin Wang,
Jun Luo,~\IEEEmembership{Fellow,~IEEE}, and
Jiming Chen,~\IEEEmembership{Fellow,~IEEE}

\thanks{Kaiyi Chi, Qianqian Yang, and Yuanchao Shu are with the State Key Laboratory of Industrial Control Technology, Zhejiang University, Hangzhou 310027, China (e-mail:\{chikaiyi17, qianqianyang20, ycshu\}@zju.edu.cn).}
\thanks{Yinghui He and Jun Luo are with the College of Computing and Data Science, Nanyang Technological University, Singapore 639798 (email:\{yinghui.he, junluo\}@ntu.edu.sg).}
\thanks{Zhiqin Wang is with the School of Information and Communication Engineering, Beijing University of Posts and Telecommunications, Beijing 100876, China, and also with China Academy of Information and Communications Technology, Beijing 100191, China (e-mail: zhiqin.wang@caict.ac.cn).}
\thanks{Jiming Chen is with the State Key Laboratory of Industrial Control Technology, Zhejiang University, Hangzhou 310027, China, and also with the School of Automation, Hangzhou Dianzi University, Hangzhou 310018, China (e-mail: cjm@zju.edu.cn).}
}



\maketitle

\begin{abstract}
Deep joint source-channel coding (DeepJSCC)
has emerged as a promising paradigm for efficient and robust information transmission. However, its intrinsic characteristics also pose new security challenges, notably an increased vulnerability to eavesdropping attacks.
Existing studies on defending against eavesdropping attacks in DeepJSCC, while demonstrating certain effectiveness, often incur considerable computational overhead or introduce performance trade-offs that may adversely affect legitimate users. 
In this paper, we present \name, to the best of our knowledge, the first physical-layer defense framework for 
DeepJSCC against eavesdropping attacks, validated through over-the-air experiments using software-defined radios (SDRs). Considering that existing eavesdropping attacks against DeepJSCC are limited to simulation under ideal channels, we take a step further by identifying and implementing four representative types of attacks under various configurations in orthogonal frequency-division multiplexing systems. 
These attacks are evaluated over-the-air under diverse scenarios, allowing us to comprehensively characterize the real-world threat landscape.
To mitigate these threats, \name introduces a novel preamble perturbation mechanism that modifies the preamble shared only between legitimate transceivers. To realize it, we first conduct a theoretical analysis of the perturbation’s impact on the signals intercepted by the eavesdropper. Building upon this, we develop an end-to-end perturbation optimization algorithm that significantly degrades eavesdropping performance while preserving reliable communication for legitimate users.
We prototype \name using SDRs and conduct extensive over-the-air experiments in practical scenarios.
Extensive experiments demonstrate that \name effectively mitigates eavesdropping threats while preserving reliable communication for legitimate users. In particular, \name can reduce the eavesdropper’s reconstruction performance by as much as 29~\!dB in PSNR and decrease classification accuracy by up to 91\% compared with the performance achieved by the legitimate user.

\end{abstract}

\begin{IEEEkeywords}
Joint source-channel coding, semantic communication, 
physical layer security, eavesdropping attack.
\end{IEEEkeywords}

\section{Introduction}

Recently, deep joint source-channel coding (DeepJSCC) has emerged as a new and promising cross-layer transmission mechanism for next-generation wireless networks~\cite{gunduz2022beyond, getu2025semantic, sagduyu2024will}. 
Unlike conventional communication systems that prioritize accurate bit-level transmission through separate source coding and channel coding modules, DeepJSCC directly compresses the source data into compact representations suitable for channel transmission. This compression is enabled by advanced deep neural networks, which play an important role in enhancing end-to-end communication efficiency.
Furthermore, DeepJSCC systems are trained in an end-to-end manner, with the physical channel explicitly incorporated as part of the neural network model. This integration significantly improves robustness against physical noise and allows the system to maintain satisfactory performance even under low signal-to-noise ratio (SNR) conditions~\cite{bourtsoulatze2019deep, dai2022nonlinear, gunduz2024joint}.
Building on these innovations, DeepJSCC has demonstrated superior performance in transmitting diverse data modalities, including images~\cite{huang2022toward, sun2023adaptive, tang2024contrastive}, speech~\cite{weng2021semantic, han2022semantic}, and video~\cite{tung2022deepwive, wang2022wireless}.

However, the unique characteristics of DeepJSCC not only offer notable performance advantages but also expose new security vulnerabilities~\cite{sagduyu2023semantic, guo2024survey, guo2025semantic}. 
Compared with conventional communication systems, DeepJSCC is inherently more susceptible to eavesdropping attacks. First, the transmitted feature representations carry rich semantic information, which can allow adversaries to infer sensitive content from intercepted signals~\cite{wang2024privacy}. Second, the end-to-end training paradigm of DeepJSCC not only improves robustness under low SNR conditions but also inadvertently benefits eavesdroppers, allowing them to recover meaningful information even in harsh channel environments where traditional communication systems typically fail.
Several studies have explored the feasibility of eavesdropping attacks on DeepJSCC systems~\cite{chen2023model, tang2025towards}, further highlighting these security threats.

To mitigate this vulnerability, a number of studies have explored enhancing the eavesdropping resilience of 
DeepJSCC. One line of work focuses on modifying the training objective to reduce semantic information leakage~\cite{marchioro2020adversarial, erdemir2022privacy, kalkhoran2023secure, zhang2023wireless, li2024secure}. However, these methods often rely on the assumption that the eavesdropper experiences worse channel conditions than the legitimate user, which may not always hold in practice. 
Another direction leverages encryption or covert communication techniques to ensure confidentiality~\cite{tung2023deep, meng2025secure, xu2023deep, chen2025enhancing, tang2025towards}. 
While these approaches offer strong security guarantees, they often introduce non-negligible computational overhead and can degrade the communication performance for authorized users due to the inherent trade-offs between security and efficiency.
In addition, the majority of existing methods have only been evaluated through simulations under idealized channel conditions, limiting their practical relevance in realistic wireless environments.
Despite the progress made, a comprehensive understanding of the real-world eavesdropping threats faced by DeepJSCC systems remains elusive.

\begin{figure}[!t]
    \centering
    \includegraphics[width=3.5in]{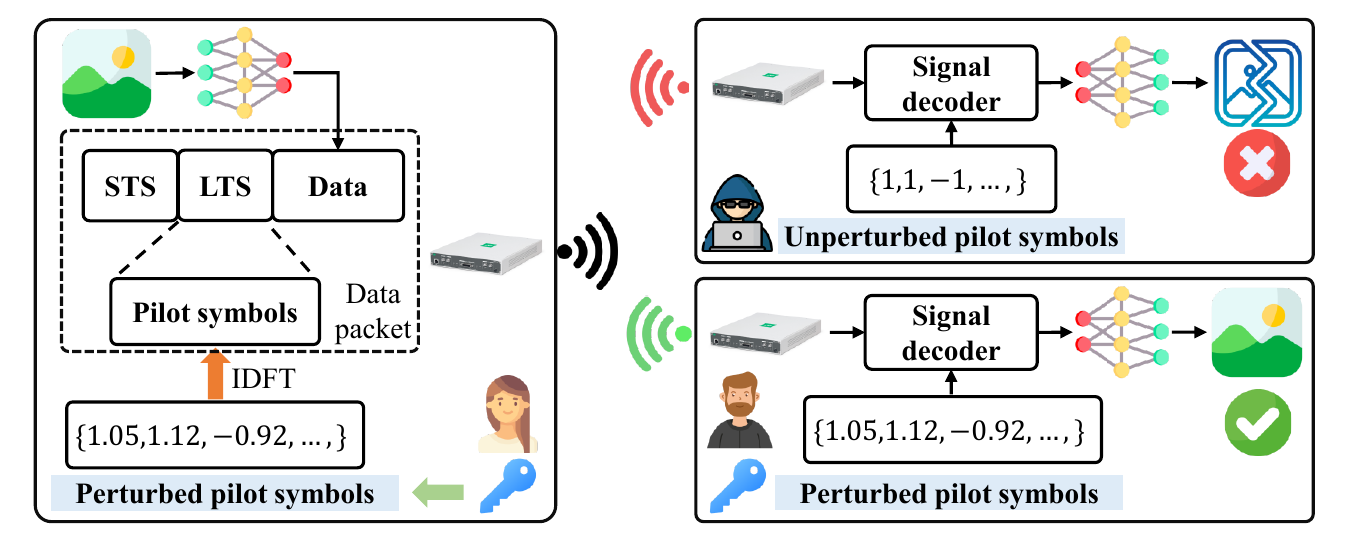}
    \caption{Illustration of the proposed \name. A perturbation is applied to the preamble between legitimate transceivers, misleading the malicious eavesdropper who attempts to decode signals using a standard preamble.}
    \label{fig:DeepGuard}
    \vspace{-1.8em}
\end{figure}

To fill this gap, we systematically identify and implement four representative eavesdropping attacks under various configurations in orthogonal frequency-division multiplexing (OFDM)-based 
DeepJSCC systems. These attacks are evaluated through over-the-air experiments using software-defined radios (SDRs), revealing the practical threats that DeepJSCC systems may face in practice. Our experimental results confirm that 
DeepJSCC is inherently vulnerable to such attacks in real-world conditions.
To counter these threats, we propose \name, which, to the best of our knowledge, is \textit{the first physical-layer defense framework for DeepJSCC systems that has been practically validated through over-the-air experiments.}
The core idea of \name is to introduce carefully crafted perturbations to the preamble, a predefined sequence commonly used for synchronization and channel estimation, and shared exclusively between legitimate transceivers. As illustrated in Fig.~\ref{fig:DeepGuard}, this mechanism misleads the eavesdropper who relies on the standard preamble to decode intercepted signals. 
We begin with a rigorous theoretical analysis of the impact of such perturbations on the signals intercepted by the eavesdropper. Building upon this, we develop an end-to-end perturbation optimization algorithm that significantly degrades the decoding performance of the eavesdropper while preserving reliable communication for legitimate users.

The proposed framework offers the following key advantages:
\textit{(i) Channel-agnostic design:} \name does not rely on assumptions about the eavesdropper’s channel quality. It remains effective across diverse scenarios, including cases where the eavesdropper has better channel conditions than the legitimate user.
\textit{(ii) Lightweight deployment:} \name requires only a few additional training epochs during the offline perturbation generation phase and \textit{introduces no extra computational overhead during the deployment phase}, as it does not involve additional signal processing modules.
\textit{(iii) Performance-preserving security:} \name offers strong protection against eavesdropping attacks without compromising the communication quality of legitimate users.
\textit{(iv) Orthogonality to existing designs:} \name can be seamlessly integrated into existing DeepJSCC workflows as a plug-in module. It can also be combined with existing defense strategies to provide enhanced security guarantees.
In summary, the major contributions of this work are as follows:
\begin{itemize}
\item We present a comprehensive study of practical eavesdropping threats in OFDM-based 
DeepJSCC system. Specifically, we identify and implement four representative attack strategies with varying configurations, covering a wide range of adversarial capabilities.

\item We evaluate these attacks through over-the-air experiments using SDRs, uncovering critical vulnerabilities that DeepJSCC systems may face in real-world environments.

\item To the best of our knowledge, we propose the first practically validated physical-layer defense framework for DeepJSCC systems that mitigates these attacks by introducing carefully designed perturbations to the preamble, which is shared exclusively between legitimate users, thereby misleading the eavesdropper.

\item We conduct a rigorous theoretical analysis to characterize the effect of preamble perturbations on the received signals at the eavesdropper. Based on this analysis, we design an end-to-end perturbation optimization algorithm
that significantly degrades the eavesdropper's performance while maintaining reliable communication for legitimate users.

\item We prototype \name on an SDR platform and conduct extensive over-the-air experiments across diverse scenarios. Experimental results confirm that \name effectively mitigates eavesdropping threats and enhances the security of 
DeepJSCC in practical deployments. In particular, \name can degrade the eavesdropper’s reconstruction performance by up to 29~\!dB in PSNR and reduce classification accuracy by up to 91\% compared with that of legitimate users.

\end{itemize}

The remainder of the paper is organized as follows. Section~\ref{sec:related_work} reviews the related work. Section~\ref{sec:preliminaries} introduces the background and threat model. Section~\ref{sec:attacks} details the considered eavesdropping attacks. Section~\ref{sec:evaluations_attack} presents the over-the-air evaluation of these attacks. Section~\ref{sec:defense} describes the proposed \name framework and Section~\ref{sec:evaluations_defense} evaluates its performance. Finally, Section~\ref{sec:conclusion} concludes the paper.

\section{Related work} \label{sec:related_work}

\subsection{Deep Joint Source-Channel Coding}
Benefiting from advances in deep learning (DL), DeepJSCC has demonstrated significant potential for improving transmission efficiency and robustness, particularly in low SNR regimes. Bourtsoulatze~\textit{et al.}~\cite{bourtsoulatze2019deep} proposed DeepJSCC for wireless image transmission, showing that DeepJSCC can effectively avoid the \textit{cliff effect} and maintain satisfactory performance under low SNR conditions. Later, Xu~\textit{et al.}~\cite{xu2021wireless} introduced an attention-based DeepJSCC architecture that improve performance across a wide range of SNRs. Tung~\textit{et al.}~\cite{tung2022deepjscc} developed DeepJSCC-Q, where the outputs of the JSCC encoder are modulated into constellation maps, improving compatibility with digital communication systems.
These approaches primarily focus on simplified additive white Gaussian noise (AWGN) channels or slow Rayleigh fading channels. Subsequent efforts have aimed to enhance applicability in more realistic settings. Yang~\textit{et al.}~\cite{yang2022ofdm} and Liu~\textit{et al.}\cite{liu2024ofdm} extended DeepJSCC to OFDM systems. Wu~\textit{et al.}~\cite{wu2024deep} further adapted it to multiple-input multiple-output (MIMO) scenarios. Bian~\textit{et al.} extended DeepJSCC to more complex cooperative relay networks~\cite{bian2025process}, 
demonstrating its versatility in diverse deployment environments.
In the deployment phase, numerous studies have explored resource allocation strategies specifically tailored for DeepJSCC networks~\cite{mu2022heterogeneous, yang2023energy, chi2024capacity, chi2024latency, liu2023adaptable}, further improving the efficiency of these systems.
While these methods have demonstrated promising results, they also introduce critical security risks. In particular, adversarial attacks and eavesdropping attacks can severely compromise the confidentiality and integrity of semantically-critical information transmitted through DeepJSCC systems.

\subsection{Adversarial Attacks on DeepJSCC}
Adversarial attacks, a type of active attack, to a DeepJSCC system, aim to compromise the transmission performance 
by injecting low-amplitude perturbations that mislead the transmitted results.
Hu~\textit{et al.}~\cite{hu2023robust} were the first to investigate adversarial perturbations in 
DeepJSCC. They examined perturbations introduced at both the transmitter and the receiver, and proposed an adversarial training strategy to mitigate their impact. Nan~\textit{et al.}~\cite{nan2023physical} proposed a method for generating input-agnostic perturbations at the physical layer to mislead classification tasks. Liu~\textit{et al.}~\cite{liu2024manipulating} developed a universal adversarial attack framework capable of deceiving classification models in both white-box and black-box scenarios. Building on prior work, Chang~\textit{et al.}~\cite{chang2023magmaw} introduced a framework that generates universal adversarial perturbations which are both modality-agnostic and protocol-agnostic. However, these studies are primarily limited to numerical simulations, while implementing adversarial attacks in real-world wireless systems remains challenging. A key obstacle is the difficulty in determining the appropriate power level of the adversarial perturbation without the knowledge of the legitimate signal’s power at the receiver side.

\subsection{Eavesdropping Attacks on DeepJSCC}
Eavesdropping attacks, which are a type of passive attack, also pose serious security threats to DeepJSCC systems. Compared with conventional communication systems, 
DeepJSCC is inherently more susceptible to such attacks due to its unique characteristics.
Chen~\textit{et al.}~\cite{chen2023model} were the first to introduce model inversion-based eavesdropping attacks~\cite{he2019model} against DeepJSCC frameworks, demonstrating that adversaries can potentially reconstruct sensitive information from the transmitted semantic features. 
Building upon this, Tang~\textit{et al.}~\cite{tang2025towards} further enhanced the attack capability by integrating generative models into the attack pipeline. 
However, these approaches are typically evaluated under idealized conditions, such as AWGN channels or Rayleigh channels with perfect channel state information (CSI) at the eavesdropper. Besides, they rely on a strong assumption that the eavesdropper has access to the raw training data of the target model, even under a black-box setting, an assumption that may not hold in realistic deployment scenarios. In addition, these methods are validated through numerical simulations, which limits their applicability and effectiveness in real-world environments with more complex and dynamic channel conditions.

Numerous efforts have been made to address eavesdropping threats in DeepJSCC systems.
A representative approach involves modifying the training procedure and loss function to enhance privacy.
For example, Zhang~\textit{et al.}~\cite{zhang2023wireless} and Li~\textit{et al.}~\cite{li2024secure} proposed loss functions that jointly optimize the reconstruction accuracy for the legitimate user and minimize semantic information leakage to the eavesdropper.
Marchioro~\textit{et al.}~\cite{marchioro2020adversarial} and Erdemir~\textit{et al.}~\cite{erdemir2022privacy} explored the theoretical underpinnings of secure DeepJSCC over wiretap channels using mutual information, and leveraged these insights to guide loss function design.
Building on this line of work, Kalkhoran~\textit{et al.}~\cite{kalkhoran2023secure} proposed a training framework and corresponding loss formulation to secure transmissions against multiple eavesdroppers.
However, these studies generally assume that the eavesdropper’s channel is worse than that of the legitimate users, which may not always hold in practice.
Moreover, there exists a fundamental trade-off between the performance of legitimate users and system security, as improving security often leads to a significant degradation in reconstruction quality for legitimate users~\cite{chen2024nearly}.

Another effective approach to enhancing security in DeepJSCC is encryption.
Tung~\textit{et al.}~\cite{tung2023deep} proposed a public-key encryption scheme tailored for 
DeepJSCC systems to defend against the eavesdropper.
Meng~\textit{et al.}~\cite{meng2025secure} explored the integration of homomorphic encryption, enabling secure computation on encoded semantic features. Rong~\textit{et al.}~\cite{rong2025semantic} proposed a physical-layer encryption scheme guided by semantic entropy.
Xu~\textit{et al.}~\cite{xu2023deep} introduced a dedicated neural network to perform visual obfuscation, aiming to preserve the privacy of transmitted images.
Chen~\textit{et al.}\cite{chen2025enhancing} applied differential privacy mechanisms to protect semantic features during facial image transmission, leveraging a computationally intensive SemanticStyleGAN framework~\cite{shi2022semanticstylegan}.
Tang~\textit{et al.}~\cite{tang2025towards} further proposed a covert communication~\cite{yang2024secure} strategy using an invertible neural network within SemanticStyleGAN to preserve semantic content in facial images.
Similar to loss-based methods, encryption-based approaches also encounter a trade-off between system security and the communication performance experienced by legitimate users.
Moreover, some of these methods require training additional neural networks to preprocess or protect private information, resulting in considerable computational overhead during the inference phase~\cite{xu2023deep, chen2025enhancing, tang2025towards}.
In addition, all these methods have so far been evaluated primarily through numerical simulations, which may limit their effectiveness in real-world environments.


\section{Preliminaries} \label{sec:preliminaries}
In this section, we first provide an overview of DeepJSCC systems, as shown in Fig.~\ref{fig:semcom}, and then formally define the threat model for eavesdropping attacks on such systems.
While we use image transmission as a representative example, the proposed attack strategies and defense framework are model-agnostic and can be readily extended to other data modalities.

\subsection{Workflow of DeepJSCC} 
\subsubsection{Transmitter}
At the transmitter, which is equipped with a single antenna, a semantic-oriented JSCC encoder  $\mathcal{E}: \mathbb{R}^N \to \mathbb{C}^M$ with parameters $\theta$ is employed to extract semantic features $\boldsymbol{X}$ from the input image $\boldsymbol{s}$:
\begin{equation}
    \boldsymbol{X} = \mathcal{E}(\boldsymbol{s}, \theta),
\end{equation}
where $N = 3 \times H \times W$ denotes the dimension of the input image, $M$ is the dimension of the encoded feature vector. The bandwidth compression ratio is defined as $M/N$.

The extracted feature $\boldsymbol{X}$ is subsequently modulated into OFDM symbols for wireless transmission. 
Specifically, $\boldsymbol{X}$ is first reshaped into a data matrix $\boldsymbol{X}' \in \mathbb{C}^{N_{\mathrm{d}} \times N_{\mathrm{s}}}$, where $N_{\mathrm{d}}$ denotes the number of active data subcarriers per OFDM symbol, and $N_{\mathrm{s}} = \lceil M / N_{\mathrm{d}} \rceil$ is the resulting number of OFDM symbols, where $\lceil\cdot \rceil$ represents the ceiling operation. Pilot symbols are inserted into predefined subcarriers, and $\boldsymbol{X}'$ is mapped onto the designated data subcarriers, while null subcarriers (e.g., DC and guard bands) remain unassigned. This yields the complete OFDM symbol matrix $\boldsymbol{X}_{\mathrm{d}} \in \mathbb{C}^{K \times N_{\mathrm{s}}}$, where $K$ is the total number of subcarriers.
This matrix is then transformed using the inverse discrete Fourier transform (IDFT), followed by cyclic prefix (CP) insertion:
\begin{equation}
    \boldsymbol{x}_{\mathrm{d}} = \boldsymbol{P} \boldsymbol{F}^H \boldsymbol{X}_{\mathrm{d}},
\end{equation}
where $\boldsymbol{F} \in \mathbb{C}^{K \times K}$ is the DFT matrix, and $\boldsymbol{P} \in \mathbb{C}^{(L_{\mathrm{cp}} + K) \times K}$ denotes the CP insertion matrix. Specifically, the CP-insertion matrix $\boldsymbol{P}$ prepends the last $L_{\mathrm{cp}}$ entries of its input vector to the beginning, i.e., 
$\boldsymbol{P} = 
\begin{bmatrix}
\boldsymbol{0} \quad \boldsymbol{I}_{L_{\mathrm{cp}}} \\
\boldsymbol{I}_K 
\end{bmatrix}$, where $\boldsymbol{I}$ denotes the identity
matrix and $\boldsymbol{0}$ denotes an all-zero matrix~\cite{he2024dual}.

\begin{figure}[!t]
    \centering
    \includegraphics[width=3.5in]{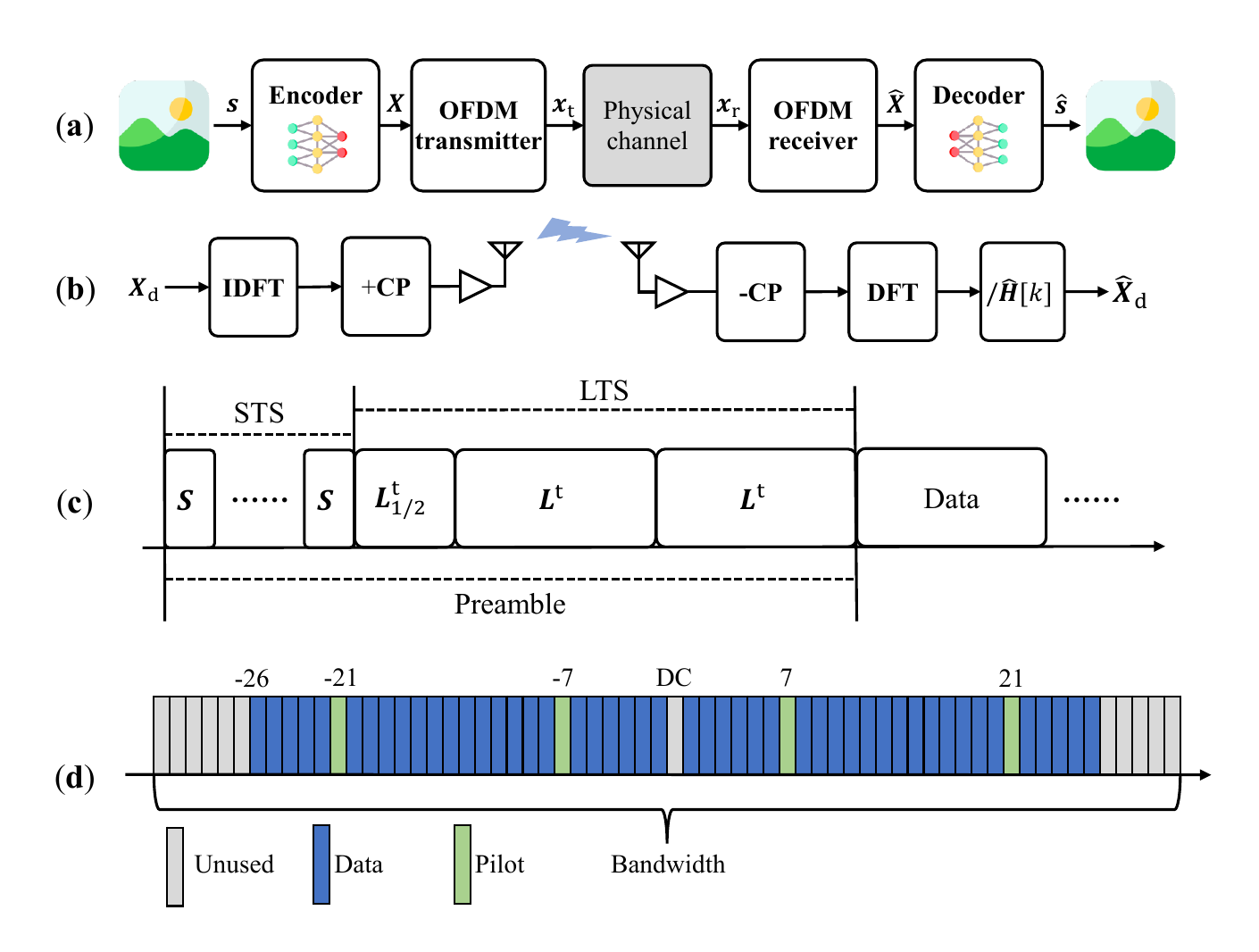}
    \vspace{-1.4em}
    \caption{(a) Workflow of DeepJSCC, (b) OFDM
transmission process of DeepJSCC, (c) structure of preamble shared between transceivers, (d) structure of subcarriers of the OFDM symbol.}
    \label{fig:semcom}
    \vspace{-0.8em}
\end{figure}

To enable synchronization and channel estimation at the receiver, a preamble sequence $\boldsymbol{x}_{\mathrm{p}}$ is prepended to $\boldsymbol{x}_{\mathrm{d}}$. As depicted in Fig.~\ref{fig:semcom}, the preamble comprises a short training sequence (STS) followed by a long training sequence (LTS). The STS consists of 10 repetitions of a 16-sample I/Q sequence (160 samples in total) and is used for packet detection, coarse timing synchronization, and coarse frequency offset correction. The LTS includes two identical time-domain OFDM symbols $\boldsymbol{L}^{\mathrm{t}} \in \mathbb{C}^{K}$ and facilitates fine timing synchronization, fine frequency offset correction, and channel estimation.
The final transmit sequence, comprising both the preamble and data symbols, is given by $\boldsymbol{x}_{\mathrm{t}} = [\boldsymbol{x}_{\mathrm{p}}, \boldsymbol{x}_{\mathrm{d}}]$ and is transmitted over the physical wireless channel.

\subsubsection{Receiver} 
The receiver, equipped with a single antenna, initiates the reception process by detecting the STS, which marks the beginning of a valid packet. The received signal is denoted as $\boldsymbol{x}_{\mathrm{r}} = [\hat{\boldsymbol{x}}_{\mathrm{p}}, \hat{\boldsymbol{x}}_{\mathrm{d}}]$, comprising the received preamble and data symbols, respectively. Upon detection, time synchronization and coarse frequency offset correction are performed using the STS.
Subsequently, the LTS is used for fine timing synchronization and fine frequency offset correction, after which the CP is removed. The receiver then applies a DFT to each of the two identical OFDM symbols in the LTS, yielding two frequency-domain vectors, denoted as $\hat{\boldsymbol{L}}_{(1)}$ and $\hat{\boldsymbol{L}}_{(2)}$. These vectors are then averaged and normalized with respect to the known frequency-domain LTS $\boldsymbol{L}$ to estimate the CSI, denoted by  $\hat{\boldsymbol{H}}[k]$, representing the estimated CSI at the $k$-th subcarrier.
Next, the CP is removed from the received data symbols, and DFT is applied to obtain the frequency-domain representation. Using the estimated CSI, equalization is performed to recover the transmitted data symbols, yielding $\hat{\boldsymbol{X}}_{\mathrm{d}}$.  Finally, the payload data are extracted from $\hat{\boldsymbol{X}}_{\mathrm{d}}$ and reshaped to reconstruct the semantic feature matrix $\hat{\boldsymbol{X}}$.

The recovered features are passed to the decoder $\mathcal{D}$ parameterized by $\phi$ to generate the final output, which may correspond to either a reconstructed image or a classification result depending on the specific transmission task:
\begin{equation}
    \hat{\boldsymbol{s}} = \mathcal{D}(\hat{\boldsymbol{X}}, \phi),
\end{equation}
Take the reconstruction task as an example, the objective is to minimize the reconstruction loss,  defined as:
\begin{equation}
    \mathcal{L}_{\mathrm{rec}} = \left\lVert \boldsymbol{s} - \hat{\boldsymbol{s}} \right\rVert_2^2, \label{eq:mse_loss}
\end{equation}
which measures the mean squared error (MSE) between the original input $\boldsymbol{s}$ and the reconstructed output $\hat{\boldsymbol{s}}$. While some studies incorporate perceptual loss to enhance visual quality~\cite{erdemir2023generative}, it is not employed in this work, though it can be seamlessly integrated if desired. To improve power amplifier efficiency, the peak-to-average power ratio (PAPR) of the transmit signal is considered, defined as:
\begin{equation}
    \rho = \frac{\max_n |\boldsymbol{x}_{\mathrm{t}}[n]|^2}{\mathbb{E}\{|\boldsymbol{x}_{\mathrm{t}}[n]|^2\}},
\end{equation}
where $\mathbb{E}\{\cdot\}$ denotes the expectation operator.
Following prior work~\cite{shao2022semantic}, we integrate the PAPR term into the training objective. The total loss function is therefore given by:
\begin{equation}
    \min_{\theta, \phi} \mathcal{L}_{\mathrm{total}} = \mathcal{L}_{\mathrm{rec}} + \lambda \mathbb{E}[\rho], \label{eq:total_loss}
\end{equation}
where $\lambda$ is a hyperparameter.
As discussed above, the direct transmission of high-level semantic features over-the-air inherently exposes the system to eavesdropping risks. In the following sections, we propose and evaluate practical attack strategies that exploit this vulnerability.

\begin{figure}[!t]
    \centering
    \includegraphics[width=3.5in]{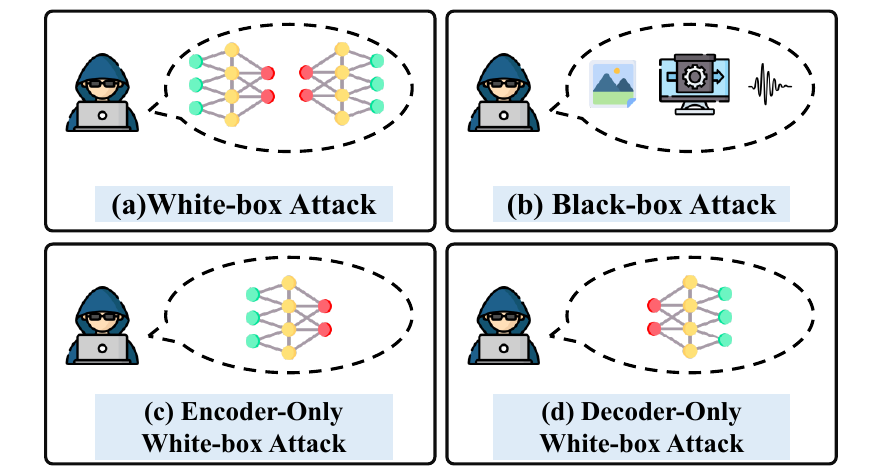}
    \vspace{-1.4em}
    \caption{Four potential eavesdropping attacks: (a) white-box attack, where the eavesdropper has complete knowledge of both the JSCC encoder and decoder; (b) black-box attack, where the eavesdropper can feed input to the device and capture the output signals; (c) encoder-only white-box attack, where the eavesdropper has complete knowledge about the JSCC encoder; (d) decoder-only white-box attack, where the eavesdropper has full knowledge about the JSCC decoder.}
    \label{fig:attack_config}
    \vspace{-0.6em}
\end{figure}

\subsection{Threat Model}
We consider a wireless communication link between a transmitter (Alice) and a receiver (Bob). By default, Alice is equipped with a JSCC encoder, while Bob uses a JSCC decoder. Both nodes are assumed to be equipped with single omnidirectional antennas. It is worth noting that the proposed framework can be readily extended to directional antennas.

An eavesdropper (Eve), also equipped with an omnidirectional antenna, passively intercepts the transmission with the goal of either reconstructing the original content or inferring the semantic output (e.g., reconstructed images or classification results), depending on the downstream task of the decoder. To achieve this, Eve employs an SDR device, such as a USRP in this work, to intercept over-the-air signals and recover encoded features. Given that the short and long training sequences (STS and LTS) are typically transmitted in the clear and follow standard formats, Eve can leverage them to perform synchronization and channel estimation for signal decoding.
Importantly, Eve does not transmit any signal and remains entirely passive, which makes the attack stealthy and difficult to detect in practical deployments.

\section{Eavesdropping Attacks} \label{sec:attacks}
In this section, we examine potential eavesdropping scenarios in practical environments and present corresponding attack strategies.

\subsection{White-box Attack}
In the white-box setting, Eve is assumed to possess complete knowledge of both the JSCC encoder and JSCC decoder, including their network architectures and parameters. This assumption is realistic in many cases, as pre-trained models are often publicly available and can be downloaded without restriction. Alternatively, if the JSCC models are trained through federated learning~\cite{nguyen2024efficient}, Eve may participate as a client and gain access to model architectures and parameter updates during the training process.

Given this full knowledge, Eve can effectively decode the intercepted signals by exploiting the publicly known STS and LTS sequences for synchronization and channel estimation. The recovered signals can then be directly processed by the known JSCC decoder to reconstruct the transmitted content or infer classification results, depending on the downstream tasks.
This setting represents a worst-case scenario from the defender’s perspective, providing an upper bound on the eavesdropping capability under ideal conditions for the attacker.

\subsection{Semi-White-box Attack}
In the semi-white-box setting, Eve possesses full knowledge of either the encoder or the decoder, but not both. This scenario can be further categorized as follows:

\subsubsection{Encoder-Only White-box Attack}
\begin{algorithm}[t]
\caption{Encoder-only White-box Attack}
\label{alg:encoder-only white-box attack algorithm}
\KwIn{
    Target encoder $\mathcal{E}(\cdot;\theta)$, auxiliary dataset $\mathcal{S}$, learning rate $\eta$, maximum number of epochs $T$.
    }
\KwOut{
    Optimized parameters $\phi^*$ of the surrogate decoder $\mathcal{D}(\cdot;\phi^*)$.
    }
Initialize surrogate decoder $\mathcal{D}(\cdot;\phi^*)$ with architecture symmetric to $\mathcal{E}(\cdot;\theta)$\;


\For{{$t = 1$ to $T$}}
    {
    \For{Each batch $\boldsymbol{s} \sim \mathcal{S}$}
        {
            Generate $\boldsymbol{\hat{s}}$ via the flow: 
        \hspace*{1.5em} $\boldsymbol{s} 
        \rightarrow \mathcal{E}(\cdot, \theta) 
        \rightarrow \text{SignalProc}_{\mathrm{Tx}} 
        \rightarrow \text{Simulated Channel} 
        \rightarrow \text{SignalProc}_{\mathrm{Rx}} 
        \rightarrow \mathcal{D}(\cdot, \phi^*)=\hat{\boldsymbol{s}}$;
        
            Compute loss $\mathcal{L}_{\mathrm{total}}$ as defined in Eq.~\eqref{eq:total_loss}\;
        
            Update decoder parameters:\newline
            \hspace*{1em} $\phi^* \leftarrow \phi^* - \eta \nabla_{\phi^*} \mathcal{L}_{\mathrm{total}}$\;
    
        \tcp{Keep encoder $\mathcal{E}(\cdot;\theta)$ fixed during training}
        }
    } 
\end{algorithm}

In this setting, Eve is assumed to have complete knowledge of the JSCC encoder. For example, she may act as a legitimate transmitter in a different 
DeepJSCC system and therefore possess full access to the encoder’s architecture and parameters. However, in the current scenario, Eve is not an authorized participant in the communication between Alice and Bob. Instead, she leverages her prior knowledge to eavesdrop on their transmission.
Additionally, we assume that Eve is aware of the general category of images transmitted by Alice. This assumption is reasonable in application-specific scenarios such as traffic monitoring, facial recognition, smart home, or visual sensing in IoT systems. Given this prior knowledge, Eve can curate an auxiliary dataset and train a surrogate model to approximate the behavior of the target JSCC decoder.

Under this setting, Eve first constructs a JSCC decoder that approximates the inverse of the encoder, exploiting the autoencoder’s inherent symmetry. She then concatenates the JSCC encoder, transmitter-side signal processing ($\text{SignalProc}_{\mathrm{Tx}}$), simulated channel, receiver-side signal processing ($\text{SignalProc}_{\mathrm{Rx}}$), and the constructed decoder into a single end-to-end network. Given her full access to the encoder, gradient backpropagation through the entire system is feasible.
Using an auxiliary dataset that shares a similar distribution with the target data, Eve can then optimize the surrogate model to effectively recover raw information. The detailed attack workflow is outlined in Algorithm~\ref{alg:encoder-only white-box attack algorithm}.

\subsubsection{Decoder-Only White-box Attack}
In this setting, Eve has complete knowledge of the JSCC decoder.
For instance, Eve may have previously participated in a different communication process that employed the same JSCC decoder, allowing her to obtain full knowledge of its design, even though she is not a legitimate receiver in the communication between Alice and Bob.
Leveraging this knowledge, Eve attempts to recover the raw content from intercepted signals transmitted between Alice and Bob.
Given full access to the decoder, Eve can directly process the received signals and perform decoding.
As a result, the attack procedure in this setting is functionally equivalent to that of the white-box scenario. In the following sections, the semi-white-box attack refers to the encoder-only white-box attack, unless stated otherwise.

\begin{algorithm}[t]
\caption{Black-box Attack}
\label{alg:black-box attack algorithm}
\KwIn{
    Target encoder $\mathcal{E}(\cdot;\theta)$, auxiliary dataset $\mathcal{S}$, learning rate $\eta$, maximum number of epochs $T$.
    }
\KwOut{
    Optimized parameters $\phi^*$ of the surrogate decoder $\mathcal{D}(\cdot;\phi^*)$.
    }
\tcp{Step 1: Initialize the surrogate decoder and build the dataset}
Initialize surrogate decoder $\mathcal{D}(\cdot;\phi^*)$\;

Construct surrogate dataset $\mathcal{S}^* = \{ (\boldsymbol{s}, \hat{\boldsymbol{X}}) \}$ via the following signal flow:\newline
\hspace*{1.5em} $\boldsymbol{s} 
\rightarrow \mathcal{E}(\cdot; \theta) 
\rightarrow \text{SignalProc}_{\mathrm{Tx}} 
\rightarrow \text{Real Channel} 
\rightarrow \text{SignalProc}_{\mathrm{Rx}} 
= \hat{\boldsymbol{X}}$\;

\tcp{Step 2: Train the surrogate decoder}
\For{{$t = 1$ to $T$}}
    {
    \For{Each batch $(\boldsymbol{s}, \hat{\boldsymbol{X}}) \sim \mathcal{S^*}$}
        {
    
        Compute decoder output:\newline
        \hspace*{1.5em} $\hat{\boldsymbol{s}} = \mathcal{D}(\hat{\boldsymbol{X}}; \phi^*)$\;
    
        Compute loss $\mathcal{L}_{\mathrm{rec}}$ as defined in Eq.~\eqref{eq:mse_loss}\;
    
        Update decoder parameters:\newline
        \hspace*{1em} $\phi^* \leftarrow \phi^* - \eta \nabla_{\phi^*} \mathcal{L}_{\mathrm{rec}}$\;
        }
    } 
\end{algorithm}

\subsection{Black-box Attack}
In the black-box setting, Eve has no direct knowledge of the JSCC encoder or decoder used in the target communication system.
However, we assume that Eve has physical access to a transmitter device equipped with the JSCC encoder, e.g., by purchasing or otherwise acquiring the hardware.
We further assume that Eve knows the general type of images being transmitted by Alice as discussed before. By feeding known images into this device and using an SDR (e.g., USRP in this work), Eve can capture the corresponding transmitted signals, thereby constructing a dataset of input-output pairs.

Under this setting, Eve aims to train a surrogate decoder that approximates the target system's output behavior. To do this, she samples batches of images from an auxiliary dataset that shares a similar distribution with the training data of the target model. Each sampled image is fed into the target transmitter, and the resulting over-the-air signal is captured using a USRP.
Eve then reconstructs the transmitted signal $\hat{\boldsymbol{X}}$ by applying the receiver-side signal processing mentioned before. This results in an auxiliary dataset $\mathcal{S} = {(\boldsymbol{s}, \hat{\boldsymbol{X}})}$, consisting of input images and their corresponding recovered feature vectors.
Since Eve has no access to the target encoder or its gradients, the encoder is excluded from training. Instead, only the surrogate decoder is optimized to minimize reconstruction loss.
The full attack procedure is summarized in Algorithm~\ref{alg:black-box attack algorithm}.

We note that for both the encoder-only white-box attack and the black-box attack, we assume that the attacker is aware of the general category of the transmitted images. This assumption inevitably enhances the attacker’s capability, effectively serving as an upper bound on the attacker’s knowledge. Therefore, if our method can successfully defend against such a strengthened attacker, it will also remain effective in scenarios where the adversary possesses less prior information.

\begin{figure}[t]
    \centering
    \setlength{\abovecaptionskip}{6pt}
    \begin{subfigure}[b]{0.95\linewidth}
        \centering
        \includegraphics[width=0.98\linewidth]{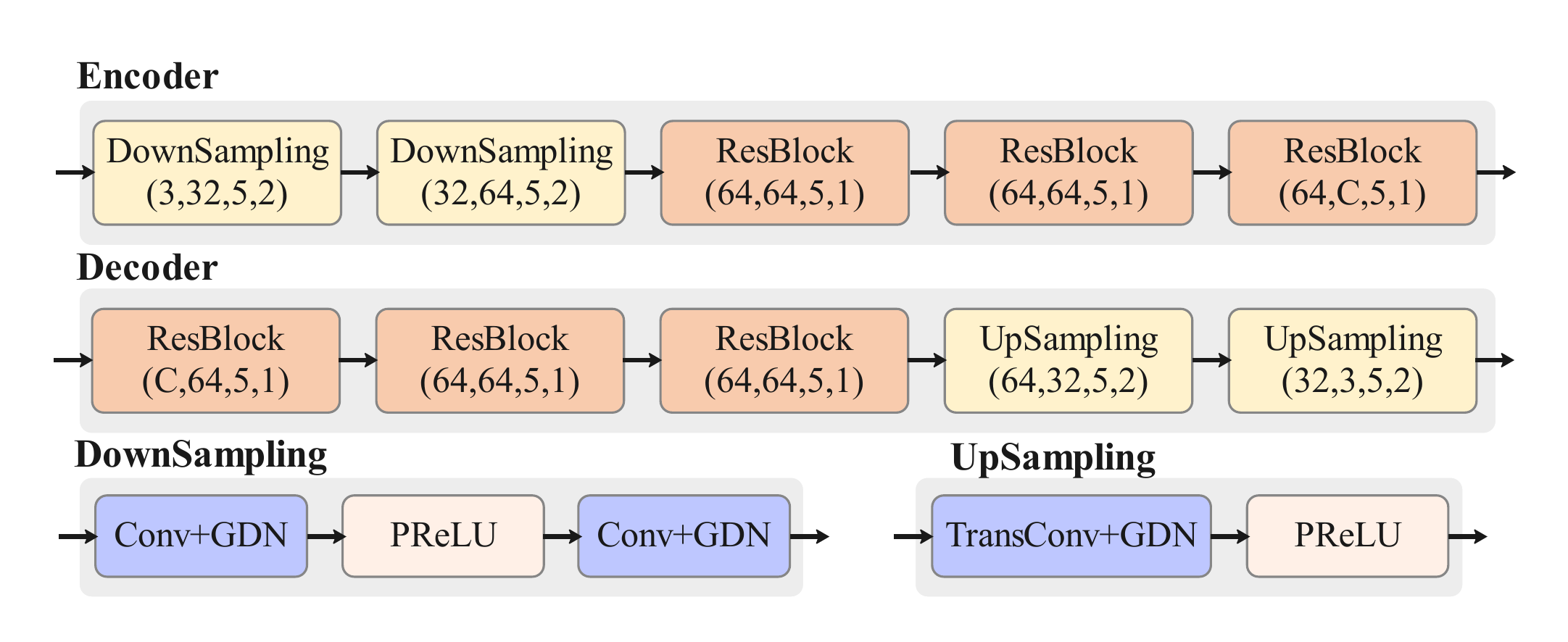}
        \vspace{-0.6em}
        \caption{Encoder and decoder for CIFAR-10 and Tiny ImageNet.}
        \label{fig:cifar_net}
    \end{subfigure}
    \begin{subfigure}[b]{0.95\linewidth}
        \centering
        \includegraphics[width=0.98\linewidth]{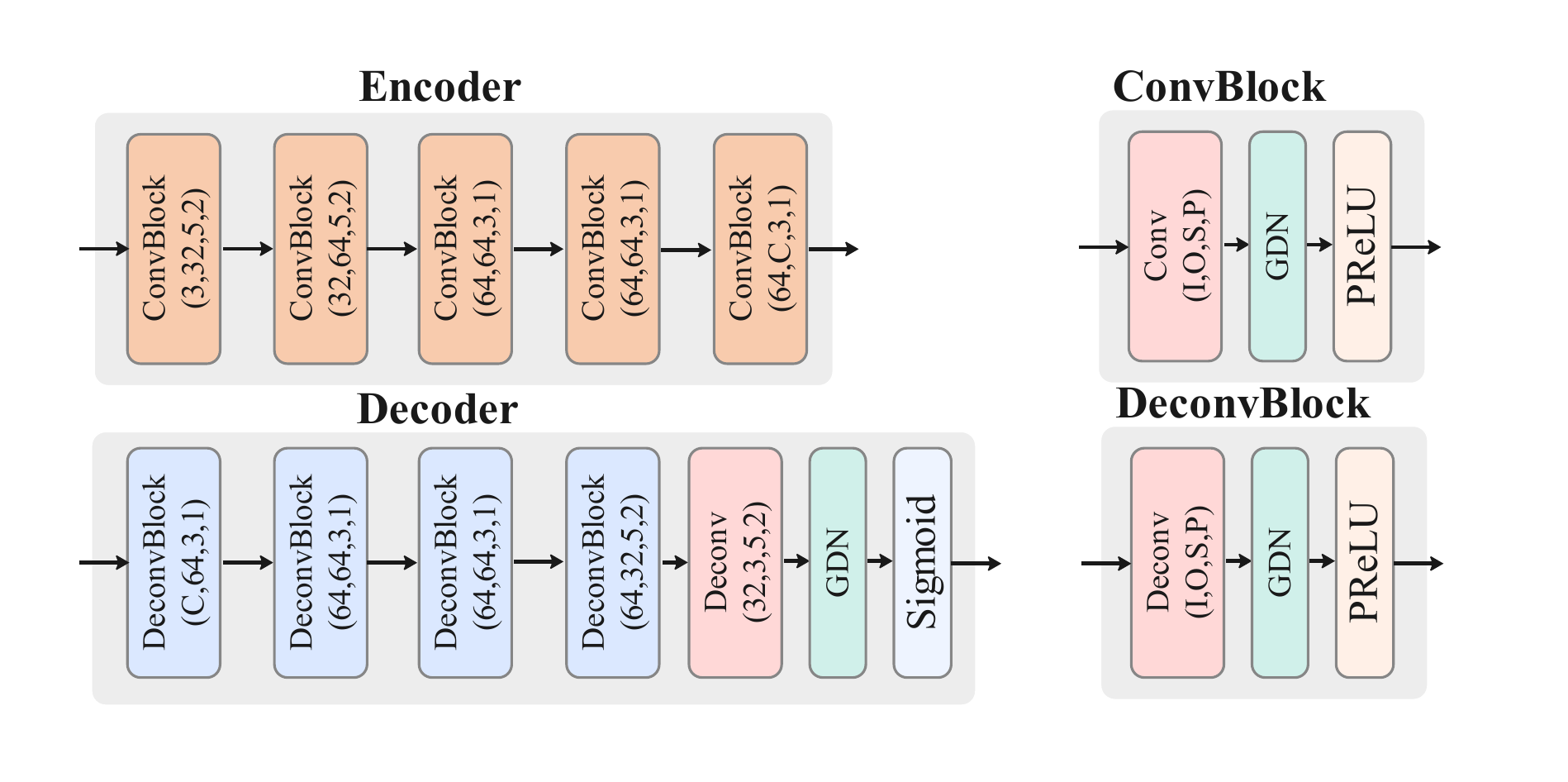}
        \vspace{-0.6em}
        \caption{Encoder and decoder for MNIST.}
        \label{fig:mnist_net}
    \end{subfigure}
    \vspace{-0.4em}
    \caption{Architecture of the neural networks, where (I, O, S, P) denotes the number of input channels, output channels, stride, and padding, respectively.}
    \vspace{-0.6em}
\end{figure}

\begin{figure}[t]
    \centering
    \setlength{\abovecaptionskip}{6pt}
    \begin{subfigure}[b]{0.45\linewidth}
        \centering
        \includegraphics[width=0.98\linewidth]{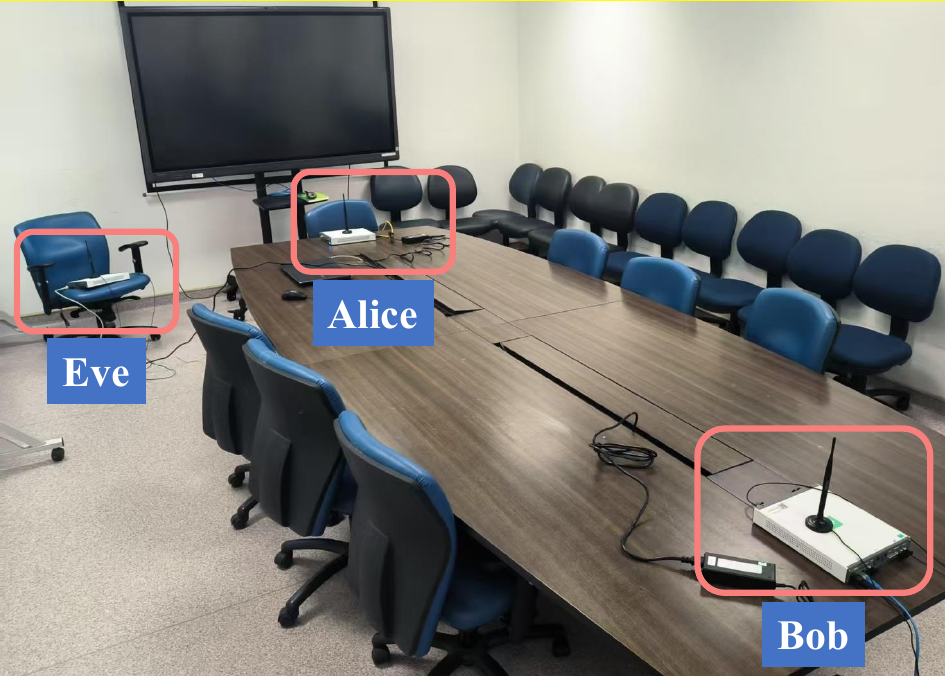}
        \vspace{-0.6em}
        \caption{LoS experiment scenario.}
        \label{fig:exp_fig}
    \end{subfigure}
    \begin{subfigure}[b]{0.48\linewidth}
        \centering
        \includegraphics[width=0.98\linewidth]{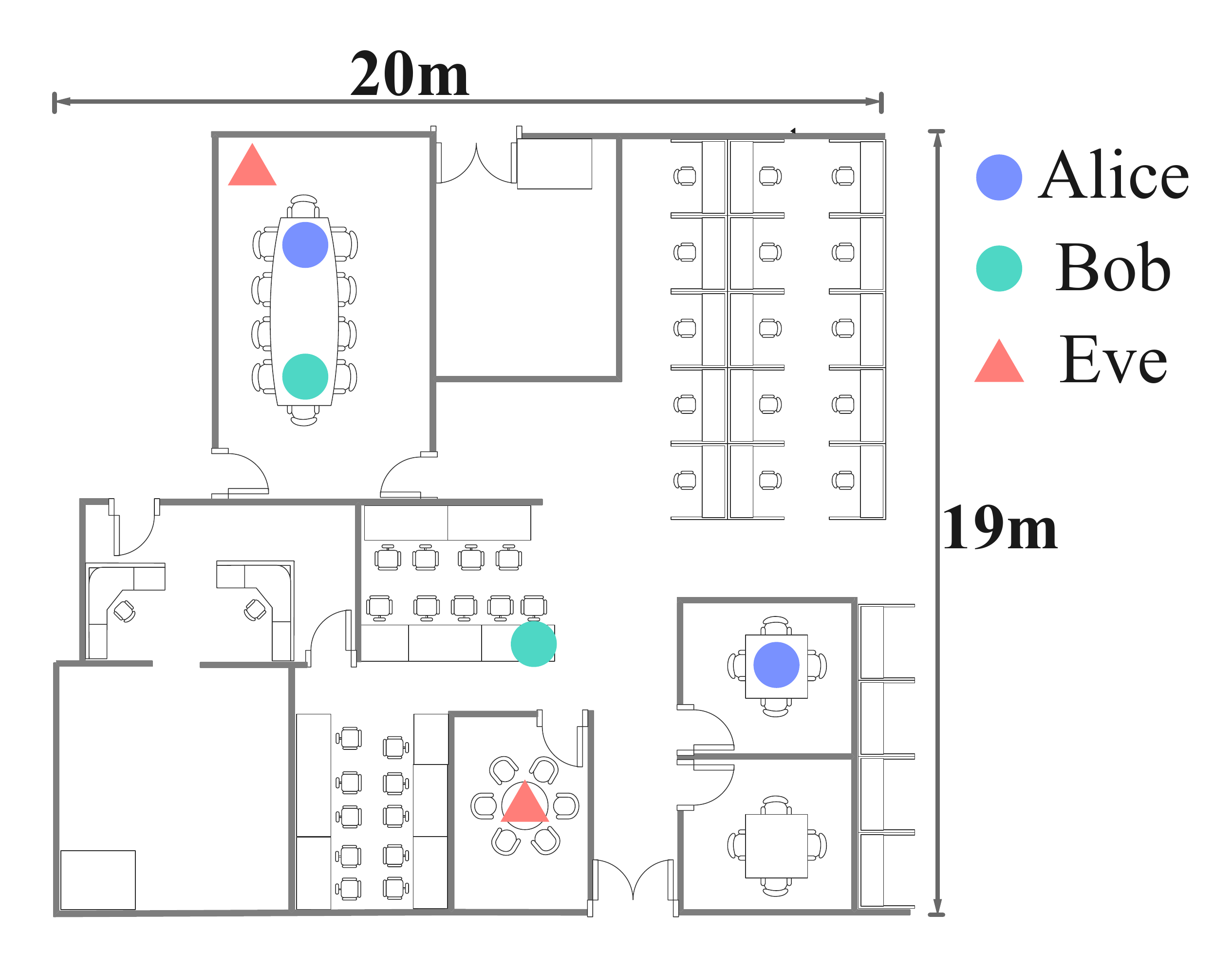}
        \vspace{-0.6em}
        \caption{Experiment layout.}
        \label{fig:scenario}
    \end{subfigure}
    \vspace{-0.4em}
    \caption{\name implementation: (a) LoS experiment scenario, and (b) experiment layout in an office area.}
    \label{fig:exp_setup}
    \vspace{-0.6em}
\end{figure}

\section{Evaluations on the Attacks} \label{sec:evaluations_attack}
In this section, we present the experimental setup and evaluate the effectiveness of the proposed eavesdropping attacks.

\subsection{Implementation and Setup}

To comprehensively evaluate the performance of the proposed framework, we conduct experiments on the CIFAR-10~\cite{krizhevsky2009learning} dataset (image size: $32\!\times\!32$), the MNIST~\cite{lecun2002gradient} dataset (image size: $28\!\times\!28$), each containing 10 classes, and the TinyImageNet~\cite{le2015tiny} dataset (image size: $64\!\times\!64$). The corresponding auxiliary datasets used for eavesdropping attacks are CIFAR-100~\cite{krizhevsky2009learning} (image size: $32\!\times\!32$), FashionMNIST~\cite{xiao2017fashion} (image size: $28\!\times\!28$), and CIFAR-100 (image size: $32\!\times\!32$), respectively.
For the CIFAR-10 and TinyImageNet datasets, the architecture of the neural network is shown in Fig.\ref{fig:cifar_net}, where $C$ is set to 16. For MNIST, the network architecture is illustrated in Fig.\ref{fig:mnist_net}, and the value of  $C$ is also set to 16. $\lambda$ is set to 1e-4 for both networks. In the black-box attack scenario, since Eve has no knowledge of the victim’s neural network, she adopts the DeepJSCC model proposed in~\cite{bourtsoulatze2019deep}, which is the most commonly used architecture, as her decoding model across all datasets.
In the classification task, a pre-trained network~\cite{chenyaofo_pytorch_cifar_models} is employed for CIFAR-10, whereas for MNIST, we adopt a simpler model, consisting of two convolutional layers followed by two fully connected layers, as the classifier.
To evaluate reconstruction quality, we use the peak signal-to-noise ratio (PSNR) as the primary performance metric. For the classification task, classification accuracy is employed as the evaluation criterion. Models are trained with PyTorch on an NVIDIA RTX A6000 GPU.

The testbed comprises three USRP X310 devices~\cite{usrpx310}: one functioning as the legitimate transmitter (Alice), one as the legitimate receiver (Bob), and the third as a passive eavesdropper (Eve). During the data collection phase for the black-box attack, Eve employs a single USRP equipped with two antennas to transmit and receive signals simultaneously.
Signal transmission and reception are controlled using the USRP Hardware Driver (UHD)~\cite{uhd}. 
All devices operate at a carrier frequency of 2.4~\!GHz and a bandwidth of 5~\!MHz. Following previous work~\cite{crow1997ieee, chi2025deepstream, 10556745}, the OFDM system employs $K = 64$ subcarriers, with $N_{\mathrm{d}} = 48$ allocated to data symbols, 4 to pilot symbols, and 12 reserved as guard bands or null subcarriers, as shown in Fig.~\ref{fig:semcom}. The detailed setting is shown in Table~\ref{tab:parameters}.

\begin{table}[t]
    \centering
    \caption{Experiment parameters}
    \label{tab:parameters}
    \begin{tabular}{lc}
    \toprule
    \textbf{Parameter} & \textbf{Value} \\
    \midrule
    Number of carriers, \( K \) & 64 \\
    Number of data carriers, \( N_{\mathrm{d}} \) & 48 \\
    Number of pilot carriers & 4 \\
    Number of null carriers & 12 \\
    CP length, \( L_{\mathrm{cp}} \) & 16 \\
    Carrier frequency  & 2.4~\!GHz \\
    Bandwidth & 5~\!MHz \\
    \bottomrule
    \end{tabular}
\end{table}

\begin{figure*}[t]
    \centering
    \setlength{\abovecaptionskip}{6pt}
    \begin{subfigure}[b]{0.32\linewidth}
        \centering
        \includegraphics[width=0.95\linewidth]{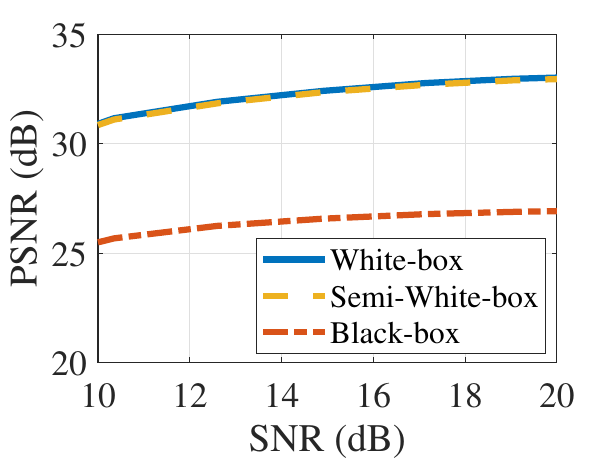}
        \vspace{-0.3em}
        \caption{CIFAR-10.}
        \label{fig:attacks_cifar_los}
    \end{subfigure}
    \begin{subfigure}[b]{0.32\linewidth}
        \centering
        \includegraphics[width=0.95\linewidth]{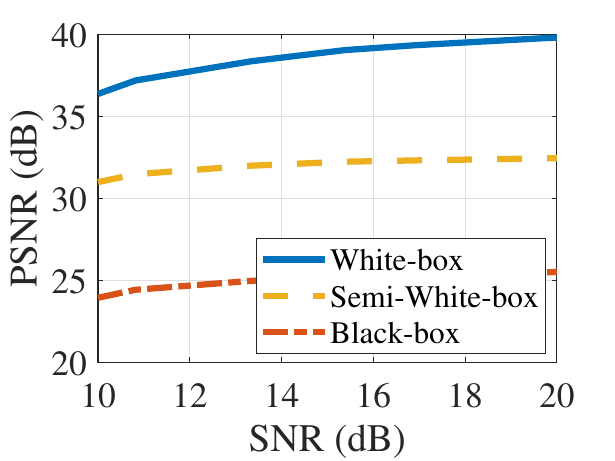}
        \vspace{-0.3em}
        \caption{MNIST.}
        \label{fig:attacks_mnist_los}
    \end{subfigure}
    \begin{subfigure}[b]{0.32\linewidth}
        \centering
        \includegraphics[width=0.95\linewidth]{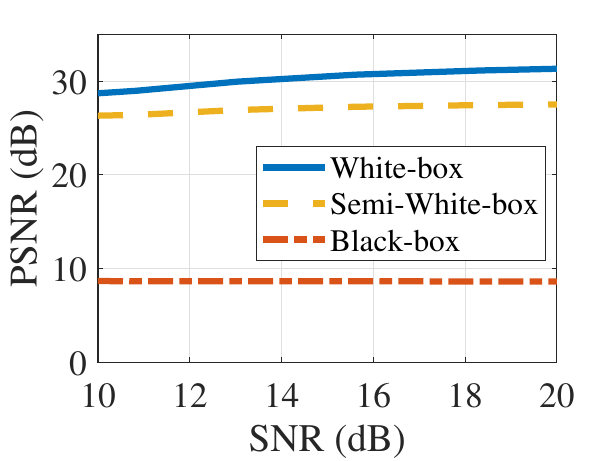} 
        \vspace{-0.3em}
        \caption{Tiny ImageNet.}
        \label{fig:attacks_imagenet_los}
    \end{subfigure}
    \vspace{-0.4em}
    \caption{Performance of the attacks over LoS scenario on (a) CIFAR-10, (b) MNIST,
    and (c) Tiny ImageNet.}
    \vspace{-0.6em}
    \label{fig:attacks_los}
\end{figure*}

We evaluate two deployment scenarios: line-of-sight (LoS) and non-line-of-sight (NLoS), as shown in Fig.~\ref{fig:exp_setup}. In the LoS setting, Eve is placed closer to Alice, resulting in better channel conditions than those experienced by Bob. In the NLoS setting, Eve is placed in a location where her channel conditions are inferior to Bob’s. We conducted over-the-air experiments for more than 40 hours, encompassing both system evaluation and the collection of datasets for black-box attacks.

\subsection{Real-World Experiment}
\subsubsection{LoS Scenario}
Fig.~\ref{fig:attacks_los} presents the performance of the proposed eavesdropping attacks across different datasets under the LoS scenario. As shown in the figure, the white-box attack consistently achieves the highest performance across all datasets, as it leverages full knowledge of the neural network. Consequently, it serves as an upper bound on the eavesdropper’s capabilities.
We observe that the semi-white-box attack exhibits performance degradation compared with the white-box attack, with its effectiveness strongly influenced by the choice of auxiliary datasets. For example, in Fig.~\ref{fig:attacks_cifar_los}, the performance of the semi-white-box attack is comparable to that of the white-box attack, whereas in Fig.~\ref{fig:attacks_mnist_los}, it suffers a significant drop relative to the white-box baseline.
Similarly, the black-box attack also demonstrates sensitivity to the auxiliary datasets. Notably, in Fig.~\ref{fig:attacks_imagenet_los}, the black-box attack experiences a substantial performance decline, primarily due to overfitting during training. 
In the following section, we will not evaluate the defense performance under this setting, as the black-box attack cannot successfully launch attacks on the Tiny ImageNet dataset.
It is worth highlighting that all three attack strategies remain effective even at an SNR of 10~\!dB, underscoring the severity of the eavesdropping threat in DeepJSCC systems.

\begin{figure*}[t]
    \centering
    \setlength{\abovecaptionskip}{6pt}
\begin{subfigure}[b]{0.32\linewidth}
        \centering
        \includegraphics[width=0.95\linewidth]{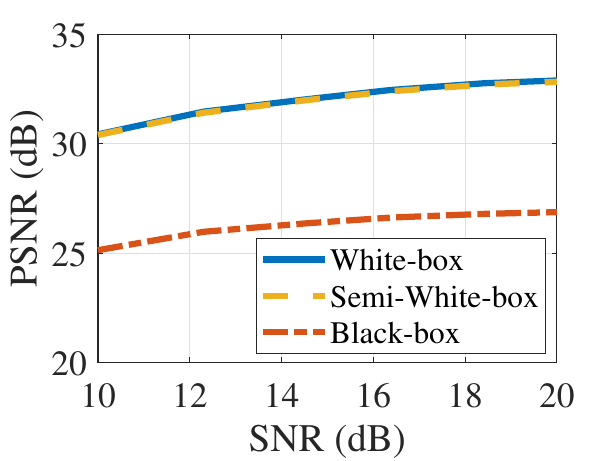}
        \vspace{-0.3em}
        \caption{CIFAR-10.}
        \label{fig:attacks_cifar_nlos}
    \end{subfigure}
    \begin{subfigure}[b]{0.32\linewidth}
        \centering
        \includegraphics[width=0.95\linewidth]{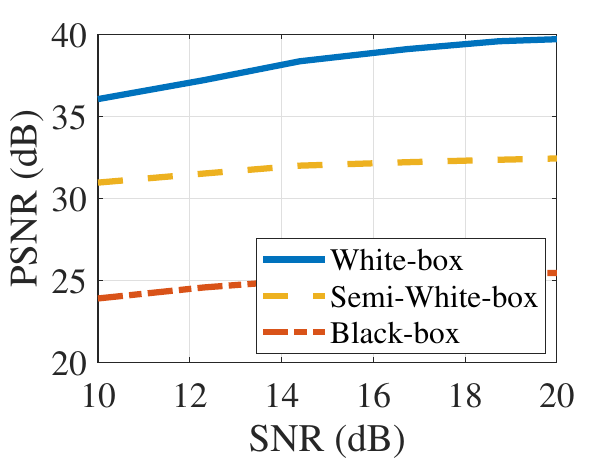}
        \vspace{-0.3em}
        \caption{MNIST.}
        \label{fig:attacks_mnist_nlos}
    \end{subfigure}
    \begin{subfigure}[b]{0.32\linewidth}
        \centering
        \includegraphics[width=0.95 \linewidth]{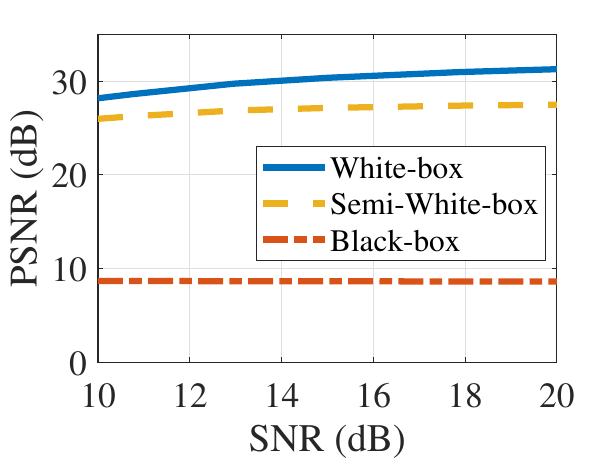}
        \vspace{-0.3em}
        \caption{Tiny ImageNet.}
        \label{fig:attacks_imagenet_nlos}
    \end{subfigure}
    \vspace{-0.4em}
    \caption{
    Performance of the attacks over NLoS scenario on (a) CIFAR-10, (b) MNIST,
    and (c) Tiny ImageNet.}
    \vspace{-0.6em}
    \label{fig:attacks_nlos}
\end{figure*}

\subsubsection{NLoS Scenario}
We then conduct eavesdropping attacks under the NLoS scenario, as shown in Fig.~\ref{fig:attacks_nlos}. The results indicate that the overall attack performance remains similar to that observed in the LoS scenario, exhibiting only slight degradation. Even under NLoS conditions, the eavesdropper is still able to extract information from the legitimate users. These findings further highlight the significant security threats posed to 
DeepJSCC systems.

\section{The design of \name} \label{sec:defense}
In this section, we introduce the proposed framework, \name. We begin by providing a theoretical analysis of how the perturbation affects the decoded signals intercepted by the eavesdropper. Subsequently, we present an end-to-end perturbation optimization algorithm that leverages this insight.

\subsection{Theoretical Analysis and Feasibility Study}
\subsubsection{Perturbation Scope in the Preamble}
To design effective perturbations, we first determine which part of the transmitted signal is suitable to be perturbed. As illustrated in Fig.~\ref{fig:semcom}, the preamble consists primarily of two components: the STS and the LTS. We analyze the roles of these components and assess their suitability for perturbation.

As discussed in Section~\ref{sec:preliminaries}, the STS consists of 10 repetitions of a 16-sample I/Q sequence. It is primarily used for packet detection, coarse timing synchronization, and coarse frequency offset correction. Due to the short length of the base sequence, the STS offers limited degrees of freedom for perturbation. Moreover, as it serves as a critical trigger for packet detection at the receiver, any modification may disrupt its autocorrelation properties and impair the ability of legitimate receivers to detect incoming packets reliably. Given these constraints, we avoid modifying the STS in our design.

In contrast, the LTS comprises two identical 64-sample symbols and is employed for fine timing synchronization, fine frequency offset correction, and channel estimation. The longer symbol length offers greater flexibility for introducing perturbations. More importantly, because LTS-based channel estimation plays a critical role in signal equalization, even minor perturbations can distort the input to the JSCC decoder, thereby reducing the eavesdropper’s decoding accuracy. Based on these observations, we focus on introducing a perturbation on the LTS. In the following subsections, we analyze how LTS modification affects the decoding performance at the eavesdropper.

As illustrated in Fig.~\ref{fig:semcom}, pilot symbols embedded within each OFDM symbol are also used to assist signal decoding, such as phase tracking. However, each OFDM symbol includes only four pilot subcarriers, providing limited degrees of freedom for perturbation. While these pilot tones offer less flexibility compared with the LTS, they could serve as a complementary target for enhancing physical-layer security when used in conjunction with LTS perturbation. In this work, we focus solely on perturbing the LTS and leave the exploration of pilot symbol perturbation as a promising direction for future research.

\begin{table}[t]
    \centering
    \caption{List of main symbols}
    \label{tab:symbol_list}
    \begin{tabular}{lc}
    \toprule
    \textbf{Symbol} & \textbf{Description} \\
    \midrule
    $\boldsymbol{x}_{\mathrm{r}}$, & Received signal \\
    $\boldsymbol{L}^\mathrm{t}$ & LTS symbol in the time domain \\
    $\boldsymbol{L}$  & LTS symbol in the frequency domain \\
    $\hat{\boldsymbol{L}}_{(1)}^{\mathrm{t}}$,  $\hat{\boldsymbol{L}}_{(2)}^{\mathrm{t}}$ & Received time-domain LTS symbols  \\
    $\hat{\boldsymbol{L}}_{(1)}$,  $\hat{\boldsymbol{L}}_{(2)}$ & Received frequency-domain LTS symbols \\
    $\hat{\boldsymbol{H}}$ & Channel estimated by Bob \\
    $\hat{\boldsymbol{H}}^{\mathrm{e}}$  & Channel estimated by Eve \\
    $\boldsymbol{W}$ & Channel noise \\
    $\boldsymbol{V}$ & Perturbation vector \\
    $\boldsymbol{A}$ & Amplitude of the perturbation vector \\
    $\boldsymbol{\varphi}$ & Phase of the perturbation vector \\
    \bottomrule
    \end{tabular}
\end{table}

\subsubsection{Impact on LTS Perturbation}
As discussed earlier, the LTS is used for fine timing synchronization, fine frequency offset correction, and channel estimation. In this subsection, we analyze how perturbing the LTS may affect each of these processing stages. The main symbols used are summarized in Table~\ref{tab:symbol_list}.

After packet detection, the receiver employs the LTS to perform fine timing synchronization by identifying the precise starting point of each OFDM symbol~\cite{10669052}. This is typically achieved via cross-correlation with the known LTS pattern. The cross-correlation at sample $i$ is computed as
\begin{equation}
    \boldsymbol{C}[i] = \sum_{j=0}^{63} \boldsymbol{x}_{\mathrm{r}}[i+j]  \overline{\boldsymbol{L}^\mathrm{t}[j]},
\end{equation}
where $\overline{\boldsymbol{L}^\mathrm{t}[j]}$ denotes the complex conjugate of the known LTS symbol in the time domain.  
While perturbing the LTS can affect the correlation values, the receiver relies only on the index of the peak cross-correlation value for timing synchronization, rather than its absolute magnitude. We will later show through experiments that such perturbations have no noticeable effect on the timing synchronization accuracy at the eavesdropper. Meanwhile, since the legitimate user is aware of the perturbed preamble, its timing synchronization remains unaffected.

Following timing synchronization, the receiver estimates and corrects the fine frequency offset using the phase difference between the two identical LTS symbols:
\begin{equation}
    \epsilon = \frac{1}{64} \angle \left( \sum_{i=0}^{63} \overline{\hat{\boldsymbol{L}}_{(1)}^{\mathrm{t}}[i]} \hat{\boldsymbol{L}}_{(2)}^{\mathrm{t}}[i] \right), \label{eq:fine_cfo}
\end{equation}
where $\hat{\boldsymbol{L}}_{(1)}^{\mathrm{t}}$ and $\hat{\boldsymbol{L}}_{(2)}^{\mathrm{t}}$ denote the first and second received time-domain LTS symbols, respectively, and $\angle(\cdot)$ represents the phase of a complex number.
The estimated offset $\epsilon$ is then used to correct each subsequent sample as
\begin{equation}
    \boldsymbol{x}_{\mathrm{r}}'[m] = \boldsymbol{x}_{\mathrm{r}}[m] e^{-j m \epsilon}, \quad m = 0,1,2,\dots
\end{equation}
Since the same perturbation is applied to both LTS symbols, their relative phase remains unchanged. As a result, fine frequency offset estimation at the eavesdropper is unaffected despite the perturbed LTS. The legitimate user, being aware of the perturbation, similarly experiences no degradation in estimation accuracy.

Finally, channel estimation is performed using the received LTS symbols to obtain the CSI, formulated as:
\begin{equation}
    \hat{\boldsymbol{H}}[k] = \frac{\hat{\boldsymbol{L}}_{(1)}[k] + \hat{\boldsymbol{L}}_{(2)}[k]}{2\boldsymbol{L}[k]}, \quad k \in \mathcal{K},
\end{equation}
where $\hat{\boldsymbol{L}}_{(1)}$ and $\hat{\boldsymbol{L}}_{(2)}$ represent the received LTS symbols in frequency domain, respectively, while $\mathcal{K}$ denotes the set of subcarriers allocated for pilots and data.
With the estimated channel, the transmitted signal on the $k$-th subcarrier of the $i$-th OFDM symbol is recovered as:
\begin{equation}
    \hat{\boldsymbol{X}}_{\mathrm{d}}[i, k] = \frac{\boldsymbol{H}[k]\boldsymbol{X}_{\mathrm{d}}[i, k] + \boldsymbol{W}[i,k]}{\hat{\boldsymbol{H}}[k]},
\end{equation}
where $\boldsymbol{W}$ represents additive white Gaussian noise.

Suppose the transmitted LTS is perturbed in the frequency domain by a complex-valued vector $\boldsymbol{V} \in \mathbb{C}^{K}$. Then, the perturbed LTS becomes:
\begin{equation}
    \boldsymbol{L}'[k] = \boldsymbol{L}[k]\boldsymbol{V}[k], \quad k \in \mathcal{K}.
    \label{eq:perturbed LTS}
\end{equation}
Accordingly, the channel estimation at the eavesdropper becomes:
\begin{equation}
    \begin{aligned}
        &\hat{\boldsymbol{H}^{\mathrm{e}}}[k] 
         = \frac{\hat{\boldsymbol{L}'}_{(1)}[k] + \hat{\boldsymbol{L}'}_{(2)}[k]}{2\boldsymbol{L}[k]}
        \\
        & = \frac{\boldsymbol{H}[k]\boldsymbol{L}'[k] + \boldsymbol{W}_{(1)}[k] + \boldsymbol{H}[k]\boldsymbol{L}'[k]+\boldsymbol{W}_{(2)}[k]}{2\boldsymbol{L}[k]}
        , \quad k \in \mathcal{K},
        \label{eq:channel_estimation at Eve}
    \end{aligned}
\end{equation}
where $\boldsymbol{W}_{(1)}$ and $\boldsymbol{W}_{(2)}$ denote the noise associated with the two LTS symbols, respectively.
When the SNR is high, the estimated channel of the eavesdropper can be approximated as 
\begin{equation}
    \hat{\boldsymbol{H}}^{\mathrm{e}}[k] = \hat{\boldsymbol{H}}[k]\boldsymbol{V}[k], \quad k \in \mathcal{K}.
\end{equation}
\textit{We can see that there is an approximate multiplicative relationship between the
estimated CSI at the legitimate user and the eavesdropper.} 
Assuming that the channel remains constant over the duration of the packet, the signal recovered by the eavesdropper is:
\begin{equation}
    \begin{aligned}
        \hat{\boldsymbol{X}}^{\mathrm{e}}_{\mathrm{d}}[i, k] 
        & = \frac{\boldsymbol{H}[k]\boldsymbol{X}_{\mathrm{d}}[i, k] + \boldsymbol{W}[i,k]}{\hat{\boldsymbol{H}}^{\mathrm{e}}[k]} \\
        & =
        \frac{\boldsymbol{H}[k]\boldsymbol{X}_{\mathrm{d}}[i, k] + \boldsymbol{W}[i,k]}{\hat{\boldsymbol{H}}[k]\boldsymbol{V}[k]}
        .
    \end{aligned}
\end{equation}
Again, when the SNR is high, the decoded signal of the eavesdropper can be approximated as
\begin{equation}
    \hat{\boldsymbol{X}}^{\mathrm{e}}_{\mathrm{d}}[i, k] = \frac{\boldsymbol{X}_{\mathrm{d}}[i, k]}{\boldsymbol{V}[k]}.
\end{equation} 
\textit{This analysis reveals that perturbing the LTS introduces subcarrier-wise amplitude and phase distortions in the eavesdropper’s decoded signal. In contrast, because the legitimate user is aware of the perturbed preamble, its channel estimation remains unaffected.} 
We experimentally validate this analytical relationship in the sequel using USRP devices.

\subsubsection{Feasibility Study}
\begin{figure}[t]
    \centering
    \setlength{\abovecaptionskip}{6pt}
    \begin{subfigure}[b]{0.48\linewidth}
        \centering
        \includegraphics[width=0.98\linewidth]{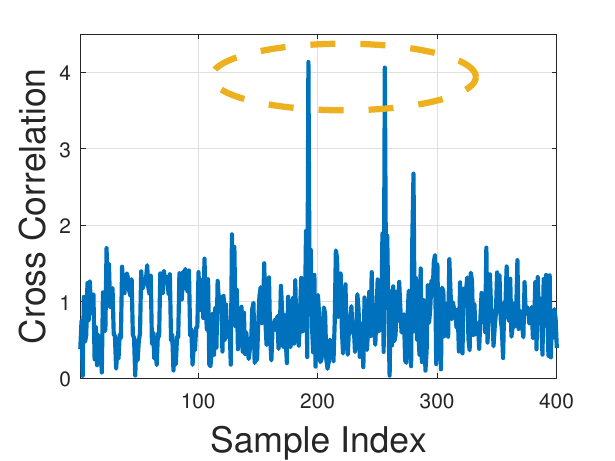}
        \vspace{-0.6em}
        \caption{With perturbed LTS.}
        \label{fig:corr_eve}
    \end{subfigure}
    \begin{subfigure}[b]{0.48\linewidth}
        \centering
        \includegraphics[width=0.98\linewidth]{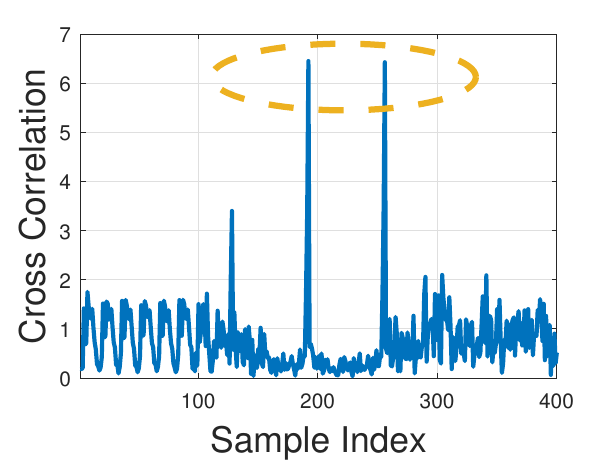}
        \vspace{-0.6em}
        \caption{Without perturbed LTS.}
        \label{fig:corr_nodefense}
    \end{subfigure}
    \vspace{-0.4em}
    \caption{Cross correlation results of Eve.}
    \label{fig:corr}
    \vspace{-0.6em}
\end{figure}

We conduct over-the-air experiments using USRPs to validate the effects of timing synchronization and channel estimation under an SNR of 20~\!dB. We first examine the impact of the perturbed LTS on timing synchronization, as illustrated in Fig.~\ref{fig:corr}. Fig.~\ref {fig:corr_eve} and Fig.~\ref{fig:corr_nodefense} show the cross-correlation results at Eve with and without the perturbed LTS, respectively. While the absolute values of the two correlation peaks decrease when the LTS is perturbed, their indices remain unchanged. Since timing synchronization relies only on the indices of the correlation peaks, the synchronization accuracy at Eve is unaffected.

\begin{figure}[t]
    \centering
    \setlength{\abovecaptionskip}{6pt}
    \begin{subfigure}[b]{0.48\linewidth}
        \centering
        \includegraphics[width=0.98\linewidth]{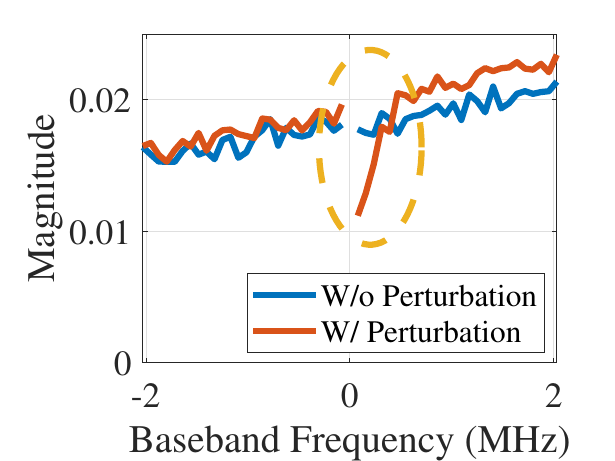}
        \vspace{-0.6em}
        \caption{CSI.}
        \label{fig:csi}
    \end{subfigure}
    \begin{subfigure}[b]{0.48\linewidth}
        \centering
        \includegraphics[width=0.98\linewidth]{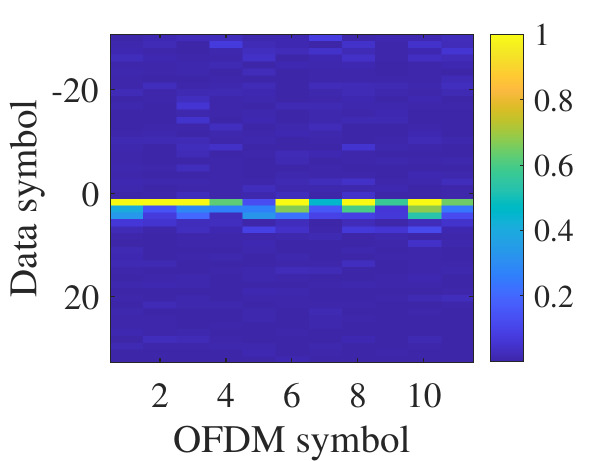}
        \vspace{-0.6em}
        \caption{MSE.}
        \label{fig:mse}
    \end{subfigure}
    \vspace{-0.4em}
    \caption{CSI and MSE of Eve.}
    \vspace{-0.6em}
\end{figure}

Next, we validate the multiplicative relationship between the perturbation and the estimated CSI. Specifically, we configure the perturbation vector as all ones, except for the 1st to 5th subcarriers following the DC subcarrier, whose values are varied from 0.5 to 0.9 in steps of 0.1. The resulting CSI estimates are shown in Fig.~\ref{fig:csi}. As expected, the absolute magnitude of the estimated CSI exhibits a linear trend consistent with the applied perturbation, confirming the theoretical multiplicative effect.
Finally, Fig.~\ref{fig:mse} presents the mean square error (MSE) of the recovered signals across subcarriers. We observe that the perturbed subcarriers exhibit significantly higher MSE, and the MSE increases as the perturbation deviates further from 1, further validating the impact of the perturbation on eavesdropper decoding.

\subsection{System Design}
In this subsection, we first introduce the perturbation generation algorithm for the reconstruction task, and then extend it to the classification task.

\subsubsection{Single Perturbation Generation}
As previously discussed, the objective of perturbation design is to mislead the eavesdropper, who relies on the standard preamble for decoding, while preserving the communication quality for the legitimate user.
As shown in the prior analysis, multiplying the LTS with a complex perturbation vector can distort the signal decoded by the eavesdropper. 

To this end, we adopt an end-to-end optimization strategy for generating the perturbation. The overall workflow is as follows. First, we parameterize the complex perturbation vector as:
\begin{equation}
    \boldsymbol{V}[k] = \boldsymbol{A}[k] e^{j\boldsymbol{\varphi}[k]}, \quad k \in \mathcal{K}, \label{eq:perturbation}
\end{equation}
where $\boldsymbol{A}$ and $\boldsymbol{\varphi}$ denote the amplitude and phase of the perturbation, respectively. Both vectors are initialized randomly.
To prevent the perturbation from degrading the performance of the legitimate receiver, we constrain $\boldsymbol{A}$ and $\boldsymbol{\varphi}$ within predefined bounds: $\alpha_{\mathrm{upper}}, \alpha_{\mathrm{lower}}, \varphi_{\mathrm{upper}}, \varphi_{\mathrm{lower}}$.
The perturbed LTS is then computed using~\eqref{eq:perturbed LTS} and inserted into the preamble along with the STS. The preamble, together with the modulated semantic features, is transmitted through a simulated wireless channel.

At the eavesdropper, channel estimation is performed using~\eqref{eq:channel_estimation at Eve}, producing a perturbed CSI. The decoded signals, distorted by this CSI, are fed into the eavesdropper’s JSCC decoder to produce a reconstructed output $\hat{\boldsymbol{s}}^{\mathrm{e}}$. 
The objective is to maximize the reconstruction error at the eavesdropper, for which we define the loss function as:
\begin{equation}
    \mathcal{L}_{\mathrm{rec}}^{\mathrm{e}} =  -\left\lVert \boldsymbol{s} - \hat{\boldsymbol{s}}^{\mathrm{e}} \right\rVert_2^2.
    \label{eq:single_loss}
\end{equation}
The amplitude and phase of the perturbation are then updated using the following gradient-based rules:
\begin{equation}
    \boldsymbol{A} = \min \left( \max \left( \boldsymbol{A} - \eta^{\mathrm{e}} \frac{ \nabla_{\boldsymbol{A}} \mathcal{L}^{\mathrm{e}}_\mathrm{rec} }{ \left\| \nabla_{\boldsymbol{A}} \mathcal{L}^{\mathrm{e}}_\mathrm{rec} \right\|_2 + \delta }, \; \alpha_{\mathrm{lower}} \right), \alpha_{\mathrm{upper}} \right),
    \label{eq:amp_update}
\end{equation}
\begin{equation}
    \boldsymbol{\varphi} = \min \left( \max \left( \boldsymbol{\varphi} - \eta^{\mathrm{e}} \frac{ \nabla_{\boldsymbol{\varphi}} \mathcal{L}^{\mathrm{e}}_\mathrm{rec} }{ \left\| \nabla_{\boldsymbol{\varphi}} \mathcal{L}^{\mathrm{e}}_\mathrm{rec} \right\|_2 + \delta }, \; \varphi_{\mathrm{lower}} \right), \varphi_{\mathrm{upper}} \right),
    \label{eq:phase_update}
\end{equation}
where $\eta^{\mathrm{e}}$ is the learning rate and $\delta = 10^{-8}$ is a small constant for numerical stability.
Since only the perturbation vectors are optimized, convergence can be achieved within a few training iterations, which will be validated in the sequel experiments. The complete procedure is summarized in Algorithm~\ref{alg:single perturbation generation}. During the generation of perturbation, we assume a white-box attack scenario, where the eavesdropper has access to the same JSCC decoder as the legitimate user. However, we will later demonstrate through experiments that the generated perturbation remains effective even under semi-white-box and black-box attacks, where the eavesdropper adopts different model parameters or architectures.

\begin{algorithm}[t]
\caption{Single Perturbation Generation}
\label{alg:single perturbation generation}
\KwIn{
    Pretrained JSCC encoder $\mathcal{E}(\cdot;\theta)$ and JSCC decoder $\mathcal{D}(\cdot;\phi)$, dataset $\mathcal{S}$, learning rate $\eta^{\mathrm{e}}$.
    }
\KwOut{
    Optimized perturbed vector $\boldsymbol{V}$.
    }
Initialize amplitude vector $\boldsymbol{A}$ and phase vector $\boldsymbol{\varphi}$\;

\For{Each batch $\boldsymbol{s} \sim \mathcal{S}$}
    {
        Generate the perturbed LTS using \eqref{eq:perturbation} and \eqref{eq:perturbed LTS}\;

        Generate $\hat{\boldsymbol{s}^{\mathrm{e}}}$ via the flow: 
        \hspace*{1.5em} $\boldsymbol{s} 
        \rightarrow \mathcal{E}(\cdot, \theta) 
        \rightarrow \text{SignalProc}_{\mathrm{Tx}} 
        \rightarrow \text{Simulated Channel} 
        \rightarrow \text{SignalProc}_{\mathrm{Rx}} 
        \rightarrow \mathcal{D}(\cdot, \phi)=\hat{\boldsymbol{s}^{\mathrm{e}}}$, where the perturbed LTS is used at the transmitter, while the standard LTS is used at the receiver\;

        Compute loss function using \eqref{eq:single_loss}\;

        Update $\boldsymbol{A}$ and  $\boldsymbol{\varphi}$ using \eqref{eq:amp_update} and \eqref{eq:phase_update}, respectively.
    }
\end{algorithm}

\subsubsection{Perturbation Codebook Generation} 
Although a single perturbation can significantly degrade the eavesdropper’s performance, relying on a fixed perturbation introduces security vulnerabilities due to its lack of diversity. To address the issue, we generate multiple perturbation vectors to construct a perturbation codebook, which is pre-shared among legitimate users prior to communication. Since the size of the codebook is small and can be encrypted using existing methods, it can be transmitted efficiently and securely. A common random seed is also shared, enabling synchronized selection of the active perturbation in each communication round. To ensure the effectiveness of the codebook, it is critical that the generated perturbations are sufficiently diverse. Otherwise, multiple perturbations may converge to similar patterns, reducing the overall robustness against eavesdropping. To promote diversity, we introduce a diversity loss term that penalizes high similarity between the current perturbation and those previously optimized.
Specifically, when generating the $J$-th perturbation, we define the diversity loss based on cosine similarity:
\begin{equation}
    \mathcal{L}^{\mathrm{e}}_{\mathrm{div}} =  \frac{1}{J-1} \sum_{i=1}^{J-1} \left( {\hat{\boldsymbol{V}}}_{J}^\top \hat{\boldsymbol{V}}_i \right)^2,
    \label{eq:diversity_loss}
\end{equation}
where $\hat{\boldsymbol{V}}_i$ denotes the normalized perturbation
\begin{equation}
    \hat{\boldsymbol{V}}_i = \frac{\tilde{\boldsymbol{V}}_i}{\|\tilde{\boldsymbol{V}}_i\|_2},
\end{equation}
and $\tilde{\boldsymbol{V}}_i$ is a concatenation of the real and imaginary parts of $\boldsymbol{V}_i$:
\begin{equation}
    \tilde{\boldsymbol{V}}_i = \begin{bmatrix} \Re(\boldsymbol{V}_i) \\ \Im(\boldsymbol{V}_i) \end{bmatrix} \in \mathbb{R}^{2K}.
\end{equation}
The total loss used for optimizing the $J$-th perturbation thus becomes:
\begin{equation}
    \mathcal{L}^{\mathrm{e}}_{\mathrm{total}} = \mathcal{L}^{\mathrm{e}}_{\mathrm{rec}} + \lambda_{\mathrm{div}} \mathcal{L}^e_{\mathrm{div}},
    \label{eq:codebook_loss}
\end{equation}
where $\lambda_{\mathrm{div}}$ is a weighting hyperparameter that balances reconstruction degradation and perturbation diversity. The overall procedure for generating the perturbation codebook is summarized in Algorithm~\ref{alg:perturbation codebook generation}.

\begin{algorithm}[t]
\caption{Perturbation Codebook Generation}
\label{alg:perturbation codebook generation}
\KwIn{
    Pretrained JSCC encoder $\mathcal{E}(\cdot;\theta)$ and JSCC decoder $\mathcal{D}(\cdot;\phi)$, dataset $\mathcal{S}$, learning rate $\eta^{\mathrm{e}}$, perturbation codebook with a pretrained perturbation $\mathcal{V}=\{\boldsymbol{V}_1\}$.
    }
\KwOut{
    Optimized perturbation codebook $\mathcal{V}$ contains $J_\mathrm{total}$ perturbed vectors.
    }
\For{$J=2$ to $J_\mathrm{total}$}
{
    Initialize amplitude vector $\boldsymbol{A}_{J}$ and phase vector $\boldsymbol{\varphi}_{J}$\;
    
    \For{Each batch $\boldsymbol{s} \sim \mathcal{S}$}
    {
    
    Generate the perturbed LTS using \eqref{eq:perturbation} and \eqref{eq:perturbed LTS}\;

    Generate $\hat{\boldsymbol{s}}^{\mathrm{e}}$ via the flow: 
    \hspace*{1.5em} $\boldsymbol{s} 
    \rightarrow \mathcal{E}(\cdot, \theta) 
    \rightarrow \text{SignalProc}_{\mathrm{Tx}} 
    \rightarrow \text{Simulated  Channel} 
    \rightarrow \text{SignalProc}_{\mathrm{Rx}} 
    \rightarrow \mathcal{D}(\cdot, \phi)=\hat{\boldsymbol{s}}^{\mathrm{e}}$, where the perturbed LTS is used at the transmitter, while the standard LTS is used at the receiver\;

    Compute loss function using \eqref{eq:codebook_loss}\;

    Update $\boldsymbol{A}_{J}$ and  $\boldsymbol{\varphi}_{J}$ using \eqref{eq:amp_update} and \eqref{eq:phase_update}, respectively, with the loss in \eqref{eq:codebook_loss};
    }
    Add $\boldsymbol{V}_{J}$ to the codebook.
}
\end{algorithm}

\subsubsection{Extend to Classification Task}
In the classification task, Eve aims to recover the classification results of the transmitted data. Since the classification task employs an asymmetric architecture, Eve must have access to the JSCC decoder in order to execute the attack. Therefore, in this context, we focus exclusively on white-box attacks.
It is worth noting that \name can be seamlessly extended to classification tasks with only minor modifications to the loss function. Specifically, we propose two defense strategies: \textit{untargeted defense} and \textit{targeted defense}.

For the untargeted defense, given an input sample $\boldsymbol{s}$ with ground-truth label $y$, the objective is to generate perturbations that cause Eve to predict any incorrect class other than the true label. The corresponding loss function is defined as
\begin{equation}
    \mathcal{L}_{\mathrm{untargeted}} = \max\Bigl(0,\; Z[\boldsymbol{s}]_y \;-\; \max_{y'\neq y}\, Z[\boldsymbol{s}]_{y'}\;+\;\kappa \Bigr),
\end{equation}
where $Z(\cdot)$ denotes the classifier logits (i.e., the outputs before the softmax function), and $\kappa$ is a constant hyperparameter that controls the confidence.

For the targeted defense, the goal is to generate perturbations that mislead Eve into predicting a predetermined target label $t$. The loss function is formulated as
\begin{equation}
    \mathcal{L}_{\mathrm{targeted}} = \max\Bigl(0,\; \max_{y'\neq t}\, Z[\boldsymbol{s}]_{y'}\;-\; Z[\boldsymbol{s}]_{t}\;+\;\kappa \Bigr).
\end{equation}
In this way, regardless of the actual transmitted data, once Eve applies the standard preamble, she will consistently classify the received signal as the specified target class.

\section{Evaluations on the \name}
\label{sec:evaluations_defense}
In this section, we evaluate the performance of the proposed \name using USRPs. 
The codebook contains eight perturbations. To generate each perturbation, we train it on the training dataset for a single epoch. The step size $\eta^{\mathrm{e}}$ is set to 0.1, and $\lambda_{\mathrm{div}} $ is also set to 0.1. The amplitude is constrained within the range 
$[0.5,2]$, while the phase is bounded between 0 and 3.14. For the classification task, we set $\kappa=0$.
Unless otherwise specified, the experimental settings are consistent with those described in Section~\ref{sec:evaluations_attack}.

\begin{figure*}[t]
    \centering
    \setlength{\abovecaptionskip}{6pt}
    \begin{subfigure}[b]{0.32\linewidth}
        \centering
        \includegraphics[width=0.95\linewidth]{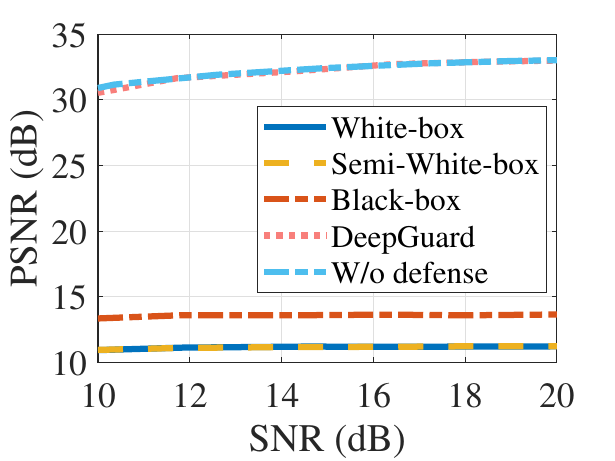}
        \vspace{-0.3em}
        \caption{CIFAR-10.}
        \label{fig:defense_cifar_los}
    \end{subfigure}
    \begin{subfigure}[b]{0.32\linewidth}
        \centering
        \includegraphics[width=0.95\linewidth]{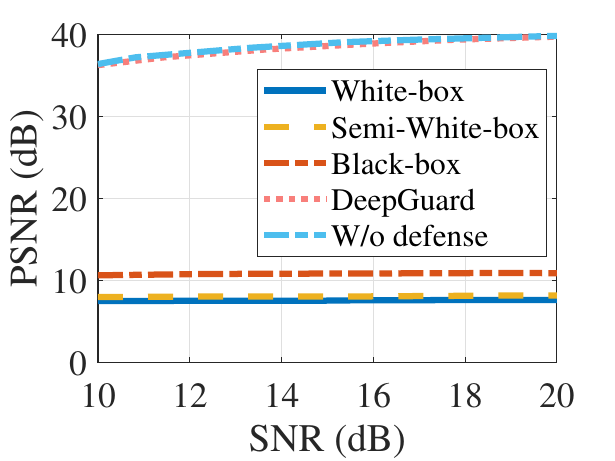}
        \vspace{-0.3em}
        \caption{MNIST.}
        \label{fig:defense_mnist_los}
    \end{subfigure}
    \begin{subfigure}[b]{0.32\linewidth}
        \centering
        \includegraphics[width=0.95\linewidth]{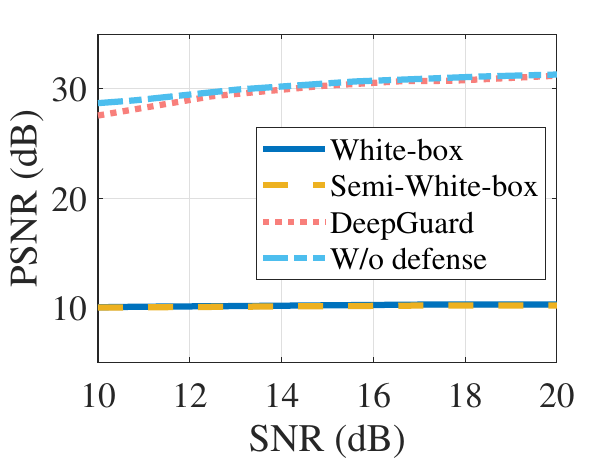} 
        \vspace{-0.3em}
        \caption{Tiny ImageNet.}
        \label{fig:defense_imagenet_los}
    \end{subfigure}
    \vspace{-0.4em}
    \caption{Performance of the \name over LoS scenario on (a) CIFAR-10, (b) MNIST,
    and (c) Tiny ImageNet.}
    \vspace{-0.6em}
    \label{fig:defense_los}
\end{figure*}

\begin{figure*}[t]
    \centering
    \setlength{\abovecaptionskip}{6pt}
    \begin{subfigure}[b]{0.32\linewidth}
        \centering
        \includegraphics[width=0.95\linewidth]{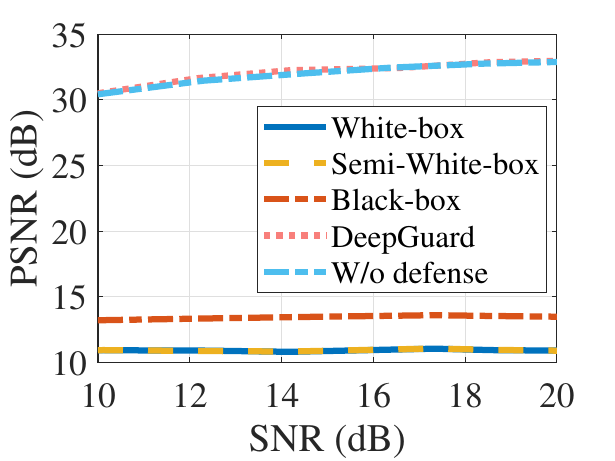}
        \vspace{-0.3em}
        \caption{CIFAR-10.}
        \label{fig:defense_cifar_nlos}
    \end{subfigure}
    \begin{subfigure}[b]{0.32\linewidth}
        \centering
        \includegraphics[width=0.95\linewidth]{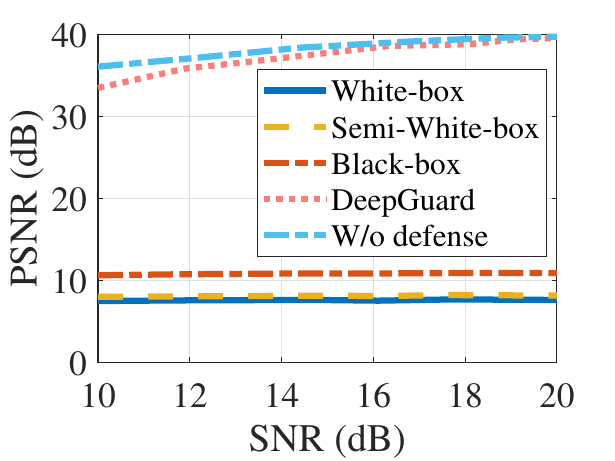}
        \vspace{-0.3em}
        \caption{MNIST.}
        \label{fig:defense_mnist_nlos}
    \end{subfigure}
    \begin{subfigure}[b]{0.32\linewidth}
        \centering
        \includegraphics[width=0.95 \linewidth]{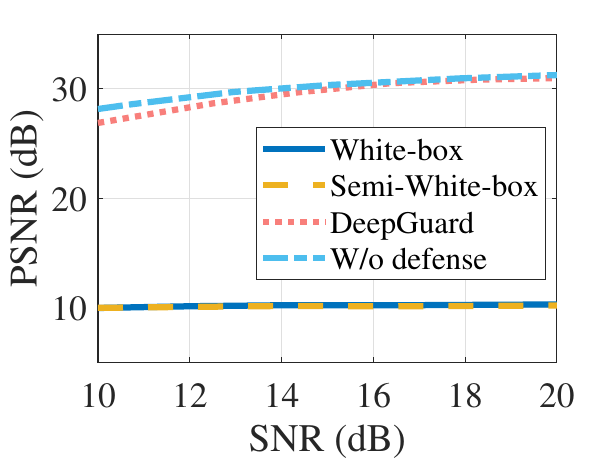}
        \vspace{-0.3em}
        \caption{Tiny ImageNet.}
        \label{fig:defense_imagenet_nlos}
    \end{subfigure}
    \vspace{-0.4em}
    \caption{
    Performance of the \name over NLoS scenario on (a) CIFAR-10, (b) MNIST,
    and (c) Tiny ImageNet.}
    \vspace{-1.6em}
    \label{fig:defense_nlos}
\end{figure*}

\begin{figure}[t]
    \centering
    \setlength{\abovecaptionskip}{6pt}
    \begin{subfigure}[b]{0.98\linewidth}
        \centering
        \includegraphics[width=0.8\linewidth]{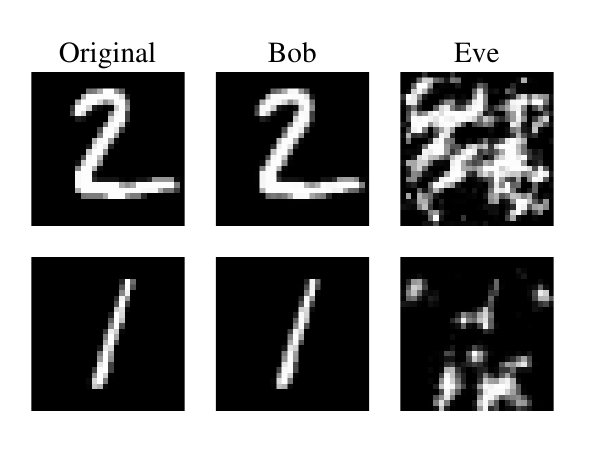}
        \vspace{-1.8em}
        \caption{MNIST.}
        \label{fig:defense_vis_mnist}
    \end{subfigure}
    \begin{subfigure}[b]{0.98\linewidth}
        \centering
        \includegraphics[width=0.8\linewidth]{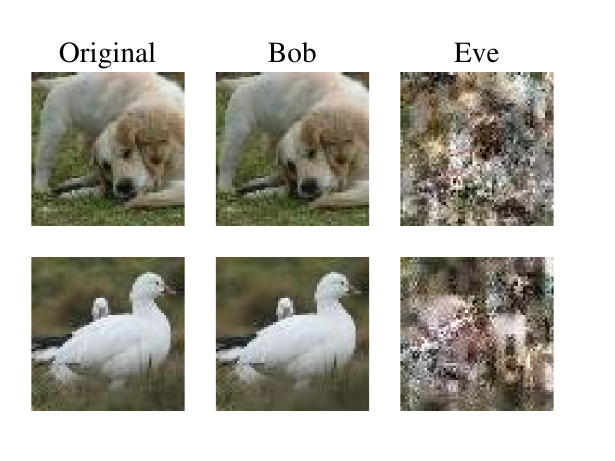}
        \vspace{-1.8em}
        \caption{Tiny ImageNet.}
        \label{fig:defense_vis_imagenet}
    \end{subfigure}
    \caption{Reconstructed images at Bob and Eve under LoS scenario at 20~\!dB SNR (white-box attack).}
    \vspace{-1.6em}
    \label{fig:defense_vis}
\end{figure}

\subsection{Overall Performance}
Fig.\ref{fig:defense_los} shows the performance of \name under the LoS scenario.
We observe that all attack strategies suffer substantial degradation in performance. Specifically, across all attacks, the reconstructed content exhibits a PSNR below 15~\!dB, indicating that Eve is unable to recover meaningful information. Notably, the white-box attack demonstrates a performance drop exceeding 20~\!dB in Fig.~\ref{fig:defense_cifar_los} and Fig.~\ref{fig:defense_mnist_los}.
Interestingly, the black-box attack achieves slightly better performance compared with the white-box and semi-white-box attacks, as it has direct access to the received signal with the modified preamble. Nevertheless, it still fails to reconstruct any semantically useful content. Meanwhile, the performance of the legitimate user remains comparable to that achieved without the proposed defense mechanism, confirming that \name effectively mitigates eavesdropping while preserving communication quality. 
As discussed earlier in Section~\ref{sec:evaluations_attack}, since the black-box attack was already unable to recover meaningful information on the Tiny ImageNet dataset, we do not evaluate it further in the defense experiments.

Fig.~\ref{fig:defense_nlos} presents the performance of \name under the NLoS scenario. Similar to the LoS case, all attack methods fail to recover any meaningful information. Meanwhile, the performance of the legitimate user remains largely unaffected, with only a slight performance drop observed on the MNIST and Tiny ImageNet datasets under low SNR conditions. These results demonstrate that \name maintains robust defense capabilities across diverse channel conditions, further validating the generalizability and practicality of the proposed framework.

Fig.~\ref{fig:defense_vis} shows the reconstructed images at an SNR of approximately 20~dB across different datasets under LoS scenario. As illustrated, Bob protected by \name can successfully recover high-quality images on both datasets. In contrast, Eve, even with a white-box attack, fails to extract any meaningful visual information. This holds for both the simpler MNIST dataset and the more complex Tiny ImageNet dataset. These results further demonstrate that \name effectively safeguards legitimate users from eavesdropping attacks while maintaining robust reconstruction performance at the legitimate receiver.

\begin{figure}[t]
    \centering
    \setlength{\abovecaptionskip}{6pt}
    \begin{subfigure}[b]{0.98\linewidth}
        \centering
        \includegraphics[width=0.67\linewidth]{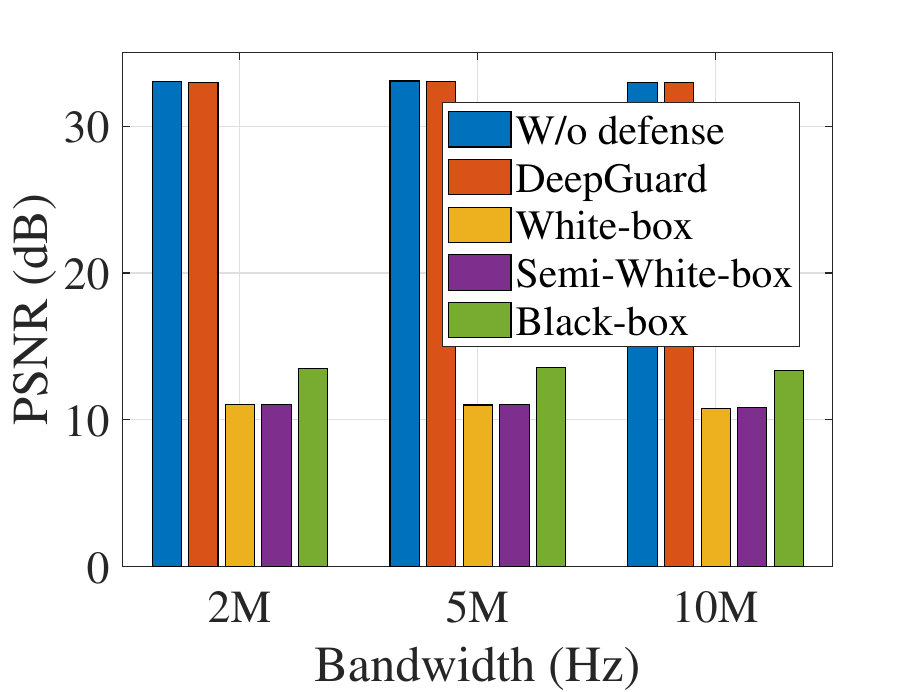}
        \vspace{-0.3em}
        \caption{Various bandwidths.}
        \label{fig:var_bw_cifar}
    \end{subfigure}
    \begin{subfigure}[b]{0.98\linewidth}
        \centering
        \includegraphics[width=0.67\linewidth]{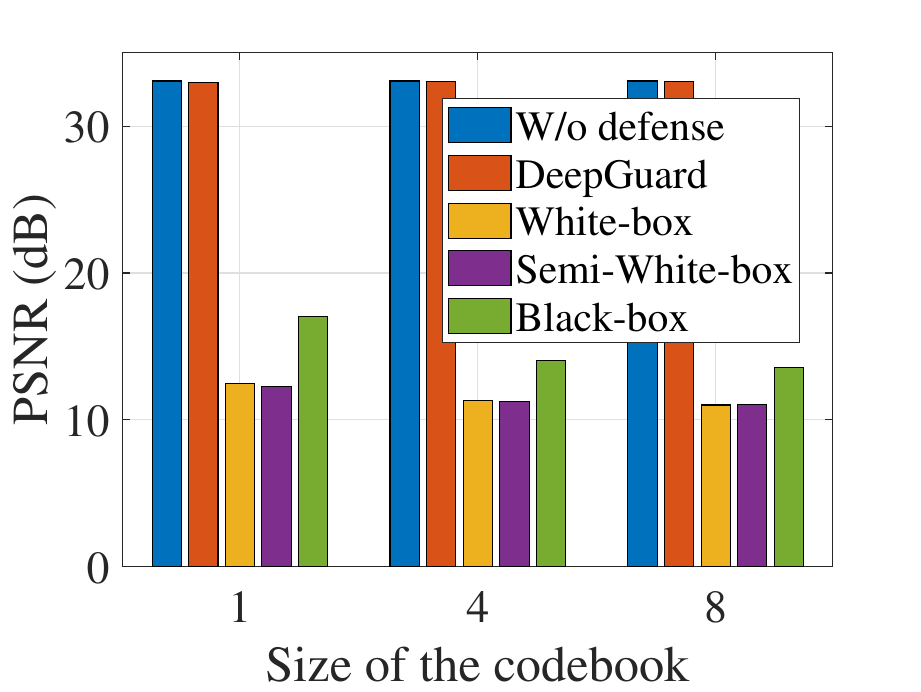}
        \vspace{-0.3em}
        \caption{Various codebook sizes.}
        \label{fig:var_code_cifar}
    \end{subfigure}
    \caption{(a) Performance under various bandwidths;  
    (b) performance under various codebook sizes.}
    \vspace{-1.6em}
    \label{fig:defense_var}
\end{figure}

\subsection{Impact Factors}
\subsubsection{Impact of Bandwidth}
We evaluate the impact of the system’s bandwidth on the performance of \name over the CIFAR-10 dataset. Specifically, we transmit signals over different bandwidths while adjusting the transmit power to maintain the SNR at 20~\!dB. Fig.\ref{fig:var_bw_cifar} shows the performance of the eavesdropper and the legitimate user across various bandwidth configurations. We observe that across all evaluated bandwidths, the performance of all attack methods remains consistently low, indicating their inability to recover any meaningful information. Meanwhile, the legitimate user consistently achieves high reconstruction quality. These results demonstrate that \name remains effective and robust across a wide range of bandwidth settings.

\subsubsection{Impact of Codebook Size}
We evaluate the impact of the codebook size on the performance of the proposed defense over the CIFAR-10 dataset. 
As shown in Fig.~\ref{fig:var_code_cifar}, increasing the codebook size leads to a reduction in the eavesdropper’s performance across all attack methods. This effect is particularly pronounced for the black-box attack. 
As discussed earlier, when the codebook contains only a single perturbation, the eavesdropper can gradually adapt and learn to compensate for its effect. In contrast, enlarging the codebook increases the diversity of perturbations, effectively preventing the eavesdropper from reliably recovering the transmitted content. Furthermore, we note that when the codebook size increases from 4 to 8, the additional performance degradation for the attacks is marginal, suggesting that a moderate codebook size is sufficient to ensure system security without incurring unnecessary complexity. This also suggests that the transmitter may randomly select perturbation vectors from a subset of the codebook rather than using the entire set, thereby introducing additional randomness and further enhancing system security.

\begin{figure*}[t]
    \centering
    \setlength{\abovecaptionskip}{6pt}
    \begin{subfigure}[b]{0.32\linewidth}
        \centering
        \includegraphics[width=0.95\linewidth]{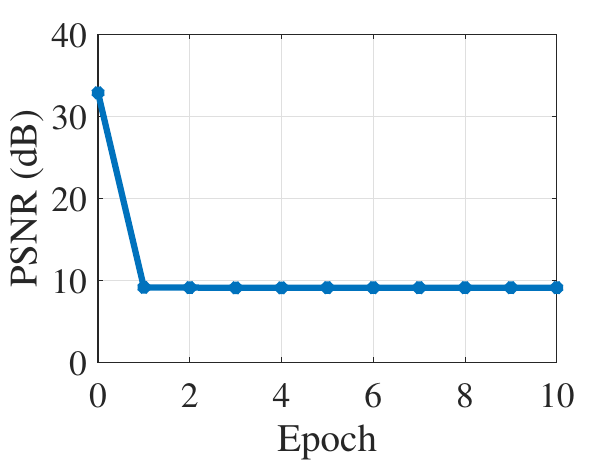}
        \vspace{-0.3em}
        \caption{Convergence (CIFAR-10).}
        \label{fig:convergence_cifar}
        \includegraphics[width=0.95\linewidth]{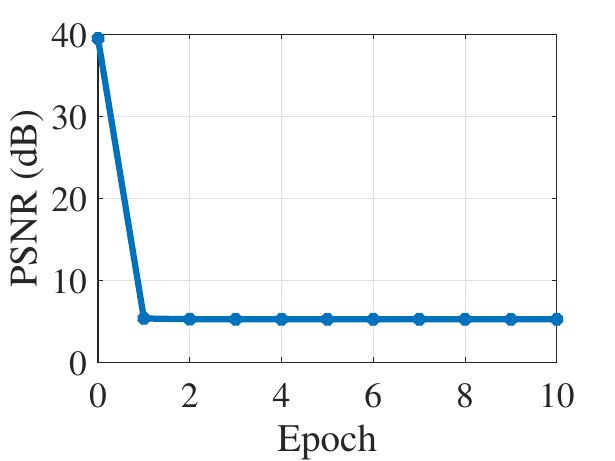}
        \vspace{-0.3em}
        \caption{Convergence (MNIST).}
        \label{fig:convergence_mnist}
    \end{subfigure}
    \hfill
    \begin{subfigure}[b]{0.32\linewidth}
        \centering
        \includegraphics[width=0.95\linewidth]{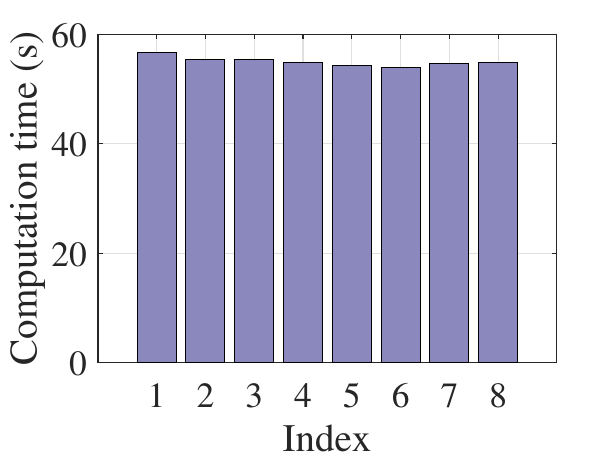}
        \vspace{-0.3em}
        \caption{Computation time (CIFAR-10).}
        \label{fig:computation_time_cifar}
        \includegraphics[width=0.95\linewidth]{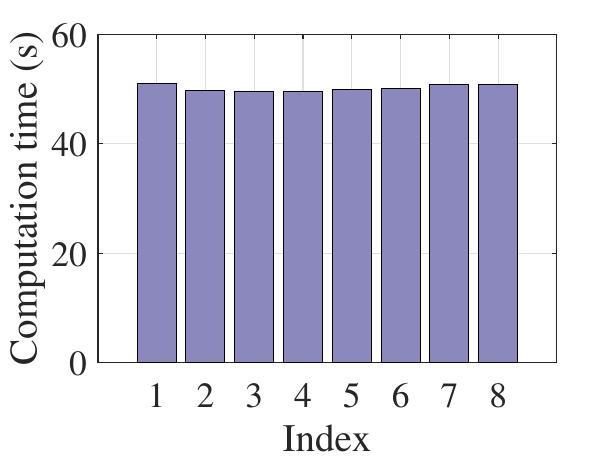}
        \vspace{-0.3em}
        \caption{Computation time (MNIST).}
        \label{fig:computation_time_mnist}
    \end{subfigure}
    \hfill
    \begin{subfigure}[b]{0.32\linewidth}
        \centering
        \includegraphics[width=0.95\linewidth]{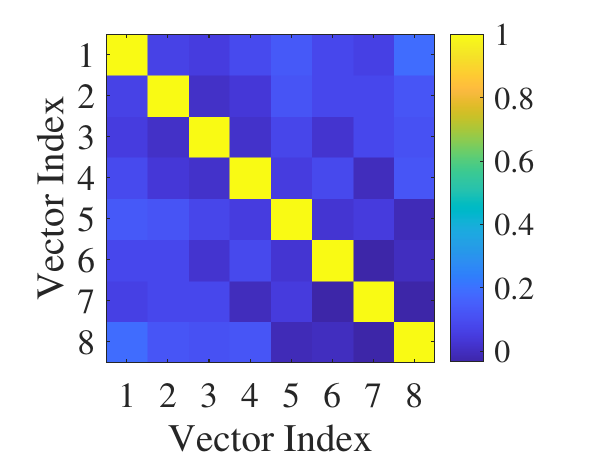}
        \vspace{-0.3em}
        \caption{Codebook diversity (CIFAR-10).}
        \label{fig:defense_cos_cifar}
        \includegraphics[width=0.95\linewidth]{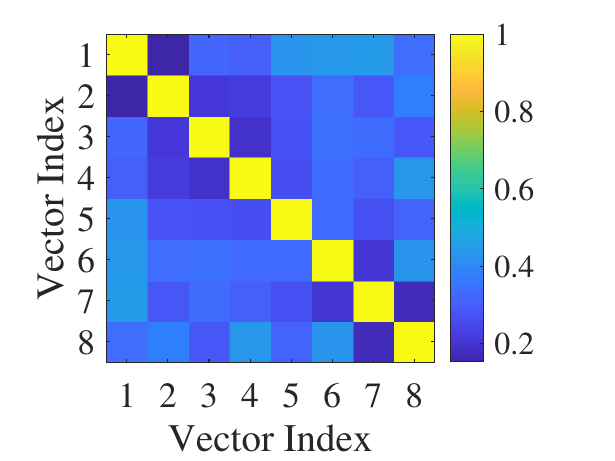}
        \vspace{-0.3em}
        \caption{Codebook diversity (MNIST).}
        \label{fig:defense_cos_mnist}
    \end{subfigure}
    \vspace{-0.4em}
    \caption{ 
    Convergence of perturbation optimization on (a) CIFAR-10, (b) MNIST;  
    computation time for codebook generation on (c) CIFAR-10, (d) MNIST;
    cosine similarity among perturbation vectors in the codebook for (e) CIFAR-10, (f) MNIST.}
    \vspace{-0.6em}
\end{figure*}

\subsubsection{Convergence Analysis}
We next evaluate the convergence speed of the perturbation optimization process. Fig.~\ref{fig:convergence_cifar} and Fig.~\ref{fig:convergence_mnist} illustrate the convergence behavior across the datasets. We observe that, for both datasets, the perturbations rapidly converge within a single training epoch once the neural network is specified. This result indicates that the computational cost of generating the perturbation codebook is minimal, enabling fast preparation of the defense mechanism in practice. Furthermore, \name does not require any additional signal processing modules at either the transmitter or the receiver, and thus introduces no extra computational overhead during deployment.

Fig.~\ref{fig:computation_time_cifar} and Fig.~\ref{fig:computation_time_mnist} show the computation time for generating the perturbation codebook on an NVIDIA RTX A6000 GPU. Each perturbation is trained for one epoch on the training dataset, and we observe that the optimization of each vector completes within one minute and remains stable. As the codebook size increases, the computation time associated with the cosine similarity loss also stays nearly unchanged. Moreover, because we currently train each perturbation for a full epoch, the computation time can be further reduced by stopping early once the loss converges. These results demonstrate that the perturbations can be optimized efficiently, enabling fast and practical deployment.

\subsubsection{Codebook Diversity} Fig.~\ref{fig:defense_cos_cifar} and Fig.~\ref{fig:defense_cos_mnist} present the cosine similarity among the perturbation vectors. We observe that the similarity between any two different vectors varies significantly. This demonstrates that the proposed training strategy effectively reduces inter-vector similarity, increases the randomness of the perturbations, and consequently enhances the overall security of the system.

\begin{figure}[t]
    \centering
    \setlength{\abovecaptionskip}{6pt}
    \begin{subfigure}[b]{0.98\linewidth}
        \centering
        \includegraphics[width=0.67\linewidth]{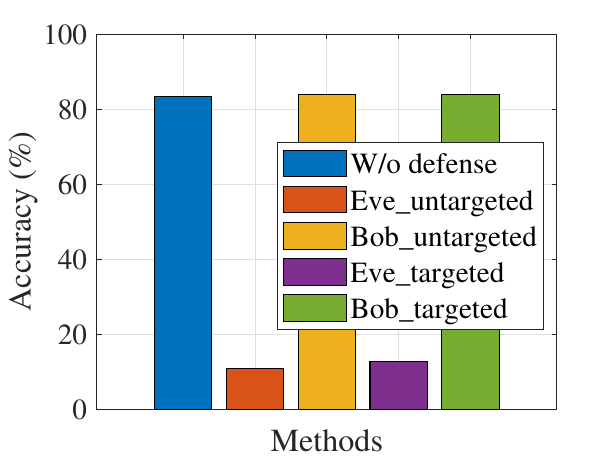}
        \vspace{-0.3em}
        \caption{CIFAR-10.}
        \label{fig:defense_cls_los_cifar}
    \end{subfigure}
    \begin{subfigure}[b]{0.98\linewidth}
        \centering
        \includegraphics[width=0.67\linewidth]{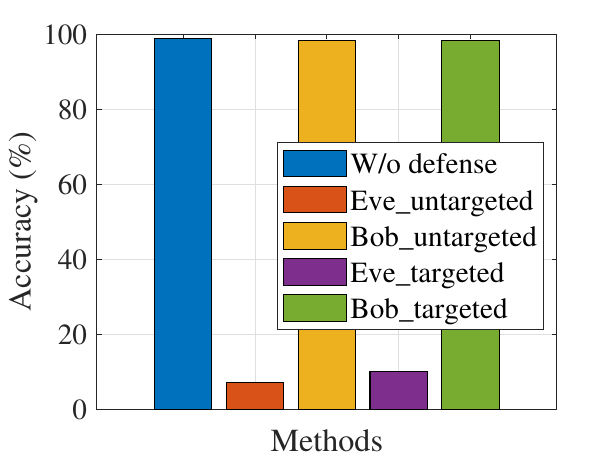}
        \vspace{-0.3em}
        \caption{MNIST.}
        \label{fig:defense_cls_los_mnist}
    \end{subfigure}
    \caption{Performance of \name for the classification task on (a) CIFAR-10, (b) MNIST.}
    \vspace{-0.6em}
    \label{fig:defense_cls}
\end{figure}

\subsection{Extend to Classification Task}
Fig.~\ref{fig:defense_cls_los_cifar} and Fig.~\ref{fig:defense_cls_los_mnist} show the performance of the eavesdropper and the legitimate user on the classification task under the LoS scenario. For both datasets, the eavesdropper’s accuracy drops to approximately 10\%, corresponding to random guessing among the 10 classes. Additionally, we observe that the untargeted defense yields slightly lower eavesdropper accuracy compared with the targeted defense, indicating stronger protection. This is because the untargeted defense strategy aims to mislead the eavesdropper into predicting any class other than the correct label, offering greater flexibility in disrupting classification. Specifically, for the MNIST dataset, the untargeted defense reduces the eavesdropper’s accuracy to as low as 7.2\%, while the classification accuracy of the legitimate user is 98.4\%, corresponding to a more than 90\% decrease compared with the baseline.
Nevertheless, we emphasize that in both defense strategies, the eavesdropper is unable to classify the received signals correctly. Meanwhile, the legitimate user consistently achieves accuracy comparable to that of the system without defense, demonstrating that \name effectively mitigates eavesdropping threats while preserving reliable classification performance.

\section{Discussion} \label{sec:discussion}

\textbf{Advanced model architectures}. In this paper, we evaluate eavesdropping attacks using CNN-based DeepJSCC models. Recent studies have investigated more advanced architectures, such as vision transformers~\cite{wu2024deep} and diffusion models~\cite{guo2025diffusion}, to further enhance the performance. Our proposed attack strategies can be naturally extended to these architectures. However, training surrogate models for such high-capacity networks typically requires larger datasets and greater computational resources, and may be even more vulnerable to overfitting, especially in black-box settings. On the defense side, the proposed framework remains broadly applicable to these advanced models, as it does not rely on any assumptions regarding the specific neural network architecture employed.

\textbf{Extension to other systems}. The OFDM system adopted in this work follows the configuration used in prior studies~\cite{chi2025deepstream}. Nonetheless, the proposed framework can be extended to cellular communication systems, as modern cellular standards also rely on known pilot signals for channel estimation, such as the Demodulation Reference Signal (DMRS) in 5G. Exploring such extensions represents an interesting direction for future work.

\section{Conclusion} \label{sec:conclusion}
In this paper, we first conducted a comprehensive study of eavesdropping threats in DeepJSCC systems and evaluated these attacks over-the-air using SDRs, demonstrating their impact in real-world environments. To address these challenges, we prototyped \name, a lightweight physical-layer defense framework designed to mitigate eavesdropping by introducing carefully crafted preamble perturbations. We performed rigorous theoretical analysis and developed an end-to-end perturbation optimization algorithm. Extensive over-the-air experiments and evaluations demonstrate that \name can effectively counter eavesdropping threats while preserving the communication performance of legitimate users.



\bibliographystyle{IEEEtran}
\bibliography{main}

\vfill

\end{document}